%% file: main.tex
\documentclass[sigconf]{acmart} 

\usepackage{array}
\usepackage{multicol,tabularx,capt-of}
\usepackage{multirow}
\usepackage{makecell}

\newcommand{\revise}[1]{{#1}} 

\AtBeginDocument{%
  }

\copyrightyear{2025}
\acmYear{2025}
\acmConference[CHI '25]{CHI Conference on Human Factors in Computing Systems}{April 26-May 1, 2025}{Yokohama, Japan}
\acmBooktitle{CHI Conference on Human Factors in Computing Systems (CHI '25), April 26-May 1, 2025, Yokohama, Japan}\acmDOI{10.1145/3706598.3713981}
\acmISBN{979-8-4007-1394-1/25/04}

\begin{document}

\title{Shape-Kit: A Design Toolkit for Crafting On-Body Expressive Haptics}

\author{Ran Zhou}
\affiliation{%
  \institution{University of Chicago}
  \city{Chicago}
  \state{Illinois}
  \country{USA}
}
\affiliation{%
  \institution{KTH Royal Institute of Technology}
  \city{Stockholm}
  \country{Sweden}
}
\email{ranzhou@uchicago.edu}

\author{Jianru Ding}
\affiliation{%
  \institution{University of Chicago}
  \city{Chicago}
  \state{Illinois}
  \country{USA}
}
\email{jrding@uchicago.edu}

\author{Chenfeng Gao}
\affiliation{%
  \institution{University of Chicago}
  \city{Chicago}
  \state{Illinois}
  \country{USA}
}
\affiliation{%
  \institution{Northwestern University}
  \city{Evanston}
  \state{Illinois}
  \country{USA}
}
\email{jessegao7@uchicago.edu}

\author{Wanli Qian}
\affiliation{%
  \institution{University of Chicago}
  \city{Chicago}
  \state{Illinois}
  \country{USA}
}
\affiliation{%
  \institution{University of Southern California}
  \city{Los Angeles}
  \state{California}
  \country{USA}
}
\email{michaelq@uchicago.edu}

\author{Benjamin Erickson}
\affiliation{%
  \institution{University of Colorado Boulder}
  \city{Boulder}
  \state{Colorado}
  \country{USA}
}
\email{beer1360@colorado.edu}

\author{Madeline Balaam}
\affiliation{%
  \institution{KTH Royal Institute of Technology}
  \city{Stockholm}
  \country{Sweden}
}
\email{balaam@kth.se}

\author{Daniel Leithinger}
\affiliation{%
  \institution{Cornell University}
  \city{Ithaca}
  \state{New York}
  \country{USA}
}
\email{dl2265@cornell.edu}

\author{Ken Nakagaki}
\affiliation{%
  \institution{University of Chicago}
  \city{Chicago}
  \state{Illinois}
  \country{USA}
}
\email{knakagaki@uchicago.edu}

\renewcommand{\shortauthors}{Zhou et al.}

\newcommand{\commentran}[1]{{\color{red}#1}} 
\newcommand{\ken}[1]{\textcolor{blue}{[KEN: #1]}}

\input{0_abstract}


\begin{teaserfigure}
  \includegraphics[width=\textwidth]{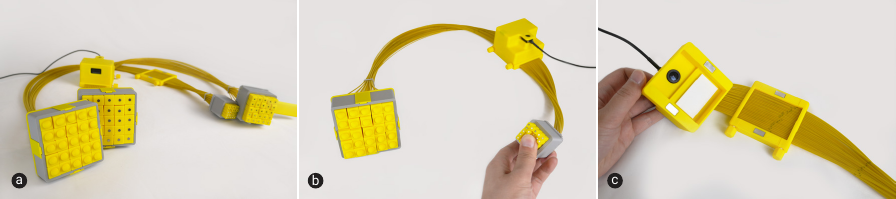}
  \caption{Shape-Kit: a hybrid haptic design toolkit for exploring crafting on-body haptics. (a) Two modules of Shape-Kit. (b) Hand behavior can be transduced to pin-based shape-change (c) Ad-hoc optical tracking module}
  \Description{3 photos of the Shape-Kit toolkit. (a) the two models analog Shape-Kit with the tracking module attached. The Shape-Kit modules are in yellow can grey color. Each module includes a pin-based shape display on both sides, connected by a cluster of bowden cables. 
 (b) A person is pressing on the small display while the pin actuation is transduced to the other side, where is the large display. (c) a close-up view of the window module on the analog Shape-Kit and the tracking module. The tracking module has a wide-angle camera and a LED light embedded
}
  \label{fig:teaser}
\end{teaserfigure}

\maketitle

\input{1_introduction}
\input{2_related_works}

\input{3_design_of_shape-kit}
\input{4_haptic_design_study}
\input{5_results_and_findings}

\input{6_takeaways_and_insights}
\input{7_discussion}
\input{8_limitations_and_future_potentials}

\input{9_conclusion}

\begin{acks}
Co-funded by the European Union (ERC, Intimate Touch, 101043637). Views and opinions expressed are however those of the author(s) only and do not necessarily reflect those of the European Union or the European Research Council. Neither the European Union nor the granting authority can be held responsible for them.

Jianru Ding’s work on this project was partially funded by Army Research Office contract number W911NF22C0082 and supported by the Intelligence Advanced Research Projects Administration (IARPA).

The authors extend their gratitude to all participants in both the pilot and final design studies. Thank you to the reviewers for their encouraging and constructive feedback. We appreciate Actuated Experience Lab, Human Computer Integration Lab, and Self-aware Computing Group members for their help with the project. We greatly thank Prof. Hank Hoffmann for his generosity in supporting Jianru Ding. We specifically thank Ramarko Bhattacharya for assisting in software development, Anup Sathya and Willa Yang for supporting hardware and study design, Prof. Harpreet Sareen for proofreading, and Emilie Faracci for photography. We are also grateful to the ATLAS Institute at the University of Colorado Boulder for their support and to Prof. Ryo Suzuki for hosting Ran Zhou in Programmable Reality Lab during the early exploration. A heartfelt thanks to Jasmine Lu for hosting Ran Zhou in Chicago and to Ran Zhou’s parents for their unwavering support.


\end{acks}

\input{main.bbl}

\input{appendix}

\end{document}

%% file: 0_abstract.tex
\begin{abstract}
Driven by the vision of everyday haptics, the HCI community is advocating for “design touch first” and investigating “how to touch well.” However, a gap remains between the exploratory nature of haptic design and technical reproducibility. We present Shape-Kit, a hybrid design toolkit embodying our “crafting haptics” metaphor, where hand touch is transduced into dynamic pin-based sensations that can be freely explored across the body. An ad-hoc tracking module captures and digitizes these patterns. Our study with 14 designers and artists demonstrates how Shape-Kit facilitates sensorial exploration for expressive haptic design. We analyze how designers collaboratively ideate, prototype, iterate, and compose touch experiences and show the subtlety and richness of touch that can be achieved through diverse crafting methods with Shape-Kit. Reflecting on the findings, our work contributes key insights into haptic toolkit design and touch design practices centered on the “crafting haptics” metaphor. We discuss in-depth how Shape-Kit’s simplicity, though remaining constrained, enables focused crafting for deeper exploration, while its collaborative nature fosters shared sense-making of touch experiences.

\end{abstract}

%% file: 1_introduction.tex
\section{Introduction}

With the maturing of haptic technology, there is a promising vision of integrating haptic interaction into everyday life, benefiting critical areas such as healthcare, communication, education, and accessibility \cite{shull2015haptic, huisman2017social, huang2010mobile, vyas2023descriptive}. Toward this vision, and given the richness of human touch perception and interpretation, the haptic community in human-computer interaction (HCI) is beginning to advocate for “design touch first” \cite{jewitt2021manifesto} and emphasizing an investigation of “how the technology can touch well” \cite{zheng2024towards}, moving beyond technology-dominant development. Approaches like Soma Design argue for designing haptic experiences center on the designer’s first-person somatic experiences \cite{hook2018embracing, hook2018designing}, which requires continuous engagement with the material on the body to judge, iterate, and validate throughout a design process \cite{balaam2024exploring, Strahl2021validity}. These methods typically employ low-fidelity prototypes or static materials (e.g., fabrics) for more visceral sensorial exploration \cite{windlin2022sketching, sondergaard2020designing}, capturing the qualitative richness of touch but limiting reproducibility. While many computational haptic design tools enable more reproducible outcomes, the burden and constraints of technical complexity often hinder more exploratory experiences for nuances. Thus, a new design approach that bridges this gap is needed.

Despite the decades of advocacy for multi-sensory interaction, touch as a design modality remains underutilized by much of the design community \cite{moussette2012simple, seifi2020novice, schneider2017haptic}. To foster design exploration in this domain, we draw inspiration from the evolution of visual design approaches. Designers have been greatly empowered by digital drawing platforms: from pixelated and later vector drawing with a mouse (e.g., Microsoft Paint \footnote{https://www.microsoft.com/en-us/windows/paint} and Adobe Illustrator \footnote{https://www.adobe.com/products/illustrator.html}) to fluid sketching with stylus pen and tablet (e.g., Procreate on iPad\footnote{https://procreate.com/}). However, the foundation of design education always starts with analog tools, such as drawing using paper or canvas with paintbrushes. They are approachable and friendly to novices, and despite their simplicity, these analog drawing methods offer versatile drawing features based on experts’ tacit knowledge, which could achieve the creation of masterpieces. In contrast, existing haptic design platforms typically consist of hardware haptic devices and software authoring tools. Most of them involve precise parameter control through sliders, buttons, line graphs \cite{schneider2016studying, zhou2023tactorbots}, or timelines with dragging blocks \cite{hapticlabs, john2024adaptics}. Some platforms use hand sketching on tablets, translating drawing patterns into actuation sequences \cite{cang2023haptic, kim2019swarmhaptics}, while some employ pressure sensors to track direct touch and replay it with a haptic interface \cite{huisman2013tasst}. However, there is a lack of haptic design tools that enable \revise{manual,} analog, and intuitive exploration, like the classical drawing on canvas with paintbrushes.

The concept of \textit{sketching haptics} emphasizes rapid prototyping with low-fi haptic design materials, focusing on experiential touch qualities \cite{moussette2012simple} and somatic appreciation \cite{windlin2022sketching}. While our work shares a similar motivation, we propose a design metaphor of “\textit{crafting haptics,}” which carries two layers of meaning. First, since haptics is inherently about touch, we aim to explore the design of expressive haptic interactions directly through hand manipulation, as intuitive as clay crafting in a dynamic way. Second, we strive to investigate how designers, using approachable, analog materials, can craft with “care, skill, and ingenuity” (Merriam-Webster \cite{mwcraft}), potentially leading to virtuoso performances in haptic design. As a first attempt, this work explores how to enable designers to “craft” dynamic pin-based force feedback that can be experienced across the body. We chose the pin display format for its potential to render versatile tactile sensations and its \revise{recognition} as a standard form factor \revise{in} haptic interfaces \cite{culbertson2018haptics}.

We present Shape-Kit, a novel hybrid haptic design toolkit that bridges the gap between exploratory design and reproducibility, embodying the “crafting haptics” metaphor. By leveraging human power and hand dexterity, the Shape-Kit analog tool can transduce and amplify (or minify) human touch behaviors into pin-based haptic sensations through a flexible and long transducer, enabling free-form sensorial exploration of touch across the body. Shape-Kit allows designers to experiment with various crafting approaches, from bare-hand manipulation to using hand-held props, similar to sculpting with clay. Textures and materials can be attached to the output end, \revise{much like} paintbrushes can have different tips for dipping in various pigments. Just as paintings can be photographed and music can be recorded, we employ an ad-hoc method to capture and digitize crafted touch patterns, which could be applied to computational pin-based haptic interfaces. Our graphical user interface (GUI) includes real-time 3D visualization, recording, tuning, and playback functionalities. To showcase a full design cycle, we built a programmable shape display for tangible playback. 
Our \textit{crafting haptics} method aims to foster intuitive analog touch prototyping, while the tracking method enables convenient digitization of the crafted outcomes. 


To characterize the “crafting” of haptics with Shape-Kit, we conducted six small group design sessions with 14 designers and artists. Inspirational topics and prompts were provided to spark the ideation, while props and materials were prepared to enhance tactile exploration. We present the design outcomes and analyze participants’ design processes through their behaviors and discussions. The results demonstrated Shape-Kit’s usability, intuitiveness, and versatility, as well as the effectiveness of its touch digitization. Our findings highlight the richness and subtle variations in touch design achieved through diverse crafting methods, from bare-hand techniques to leveraging the affordances, properties, and textures of the props and materials. We also gained insights into how designers collaboratively ideate, prototype, iterate, and compose bodily haptic experiences with Shape-Kit, typically through rapid role-switching between crafter, perceiver, and holder.

\revise{Reflecting on the findings, our work contributes key insights into haptic toolkit design and touch design practices centered on the “crafting haptics” metaphor. The analog crafting method offers an intuitive entry point for touch prototyping while excelling at uncovering subtle nuances that shape touch quality, making it well-suited for designing contextually pleasant and emotionally resonant touch experiences. Shape-Kit’s touch digitization feature supports touch recording and playback while enhancing reflective creation. We also discuss in depth how crafting haptics involves inherent constraints that can also present opportunities and how collaborative analog touch crafting fosters shared sense-making and deepens the resonance of touch experiences. Finally, we outline Shape-Kit’s current limitations, propose potential improvements, and suggest how its crafted outcomes could be applied in real-world contexts through emerging haptic technologies.}


%% file: 2_related_works.tex
\section{Related Works}
\subsection{Authoring Methods for Haptic Design Tools}
Designing haptic interactions to convey expressive and meaningful messages has gained increasing interest, driving the development of haptic design toolkits. Most haptic toolkits introduced by the HCI community consist of authoring software that controls off-the-shelf or custom hardware haptic interfaces. These toolkits span a variety of haptic technologies, while the authoring methods typically rely on GUIs with buttons and sliders to adjust parameters such as vibration amplitude and frequency \cite{paneels2013tactiped}, servo rotation and speed \cite{zhou2022emotitactor, zhou2023tactorbots}, or fluid-based valve control \cite{kilic2021omnifiber, shtarbanov2023sleevio}. 2D animation editing methods have also been explored in haptics, including tunable line graphs \cite{ryu2008posvibeditor, schneider2016studying}, digital sketching \cite{cang2023haptic, kim2019swarmhaptics}, and drag-and-drop timelines \cite{ryu2008posvibeditor, schneider2016studying, hapticlabs}. More advanced digital simulations with direct manipulation have been applied to vibrotactile actuator arrays \cite{schineider2015tactile}, mid-air haptics \cite{seifi2023feellustrator}, and for shape-change designs involving shape memory alloy (SMA) \cite{messerschmidt2022anisma, messerschmidt2024cotacs}. Despite their precision and sophistication, these methods are still screen-based, requiring designers to alternate between visual editing and tactile feedback, which can hinder more serendipitous exploration. \revise{HapticPilot \cite{sung2024hapticpilot} introduces a more in-situ approach where the designer can sketch directly on hand to author the vibrotactile patterns for haptic gloves, while this method is constrained to the virtual reality (VR) context.}

Given the unique feature that vibration and auditory cues share the same source, researchers have introduced musical metaphors into haptic design \cite{lee2009vibrotactile}, enabling applications such as collaborative improvisation \cite{schneider2014improvising}, game effect authoring \cite{degraen2021weirding}, and converting object interaction sounds into realistic vibration signals \cite{minamizawa2012techtile}. While vibration-based interaction design is maturing, vibrotactile feedback alone is often perceived as artificial \cite{messerschmidt2024cotacs}, failing to capture the rich complexity of human touch. For instance, TassT features a more embodied method where designers can author the sleeve with a vibrotactile array simply by pressing the upper sensor layer with hands \cite{huisman2013tasst}. However, it could not recreate the physical touch that contained force feedback. Our work focuses on dynamic tactile force feedback, a more natural modality that mirrors everyday object interaction and social touch. Although there is no straightforward signal translation like \revise{audio}-to-vibration for this modality, dynamic shape-change might suit rendering these sensations. We propose that such interactions can be facilitated through “crafting-based” design methods, particularly as we aim to develop a toolkit that enables more nuanced, body-centric sensorial exploration.

\subsection{Sensorial Exploration for Touch Design}
While many of the tools mentioned above prioritize technical innovation and functionality, there is a growing interest in the design research community in sensorial exploration, particularly for creating expressive and meaningful experiences that are rich, nuanced, and often personal. Camille Moussette’s  Simple Haptics approach \cite{moussette2012simple} argues for bridging interaction design and haptics through “sketching,” with “a strong focus on experiential and directly experiencable perceptual qualities of haptics.” In body-centric design, Sensory Bodystorming uses ideation probes with diverse material qualities to generate ideas for sensory augmentations \cite{turmo2018sensory}. Meanwhile, the Soma Design approach \cite{hook2018designing} emphasizes rooting the design in felt, bodily experiences while asking the designers to keep engaging with their own soma (body, mind, and emotion) throughout the design process. Their practices are often grounded on first-person experiences, using low-fidelity, static, or everyday materials, and benefit from the manual support of close collaborators. This allows designers to focus on somaesthetic appreciation and produce delicate outcomes that deliver the right feel  \cite{jonsson2016aesthetics, tsaknaki2021breathing, strahl2022annotated}. To facilitate the Soma Design practice in various contexts, researchers developed toolkits like Soma Bits \cite{windlin2022sketching}, Menarche Bits \cite{sondergaard2020designing}, and Pneumatic Prototyping Kit \cite{balaam2024exploring} which enable sensorial engagement across the body in first-person and experience sharing with others. However, given their exploratory nature, those tools are mostly open-ended, with outcomes captured primarily in qualitative terms. There is a lack of methods to record, playback, or further fine-tune these designs as computational systems can offer. Shape-Kit addresses this gap by aligning with body-centric design motivations while introducing the first hybrid haptic design tool that combines the strengths of analog prototyping, enabling the intuitive crafting of dynamic shape sensations for serendipitous exploration with the precision of computational digitization and documentation. This balance fosters both exploratory and reproducible design practices.

\begin{figure*}[ht]
  \includegraphics[width=\textwidth]{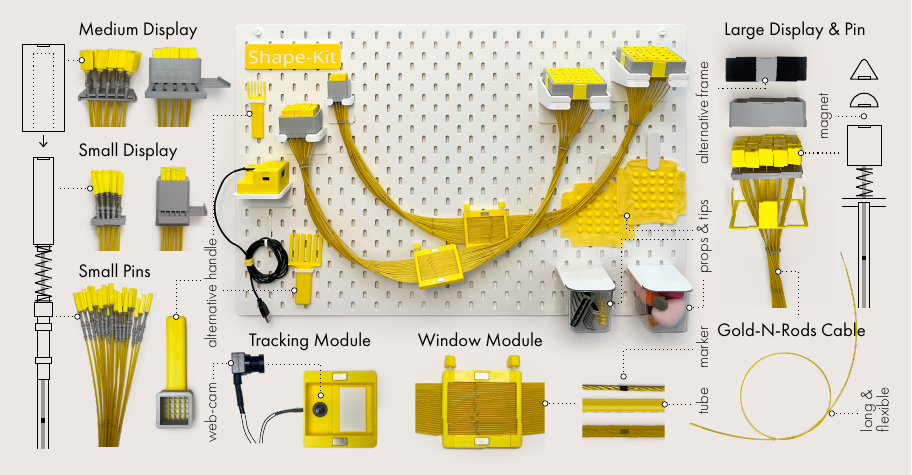}
  \caption{Shape-Kit Toolkit and its assembly}
  \Description{a figure shows the assembly of the analog Shape-Kit and the tracking module. It includes all the components for the toolkit hanging on a pegboard in the center, surrounded by the internal mechanism for all the modular parts of the Shape-Kit.}
  \label{fig:analog_system}
\end{figure*}

\subsection{Pin-based Shape Displays}
\label{sec:pin-based}
To craft diverse tactile force feedback, we identified dynamic shape-changing via pin-based shape displays as a suitable format. By altering the physical shape of an array of pins, these displays can create versatile, 2.5-dimensional animated forms that generate a wide range of sensations. Shape displays have been explored at various scales. Large-scale tabletop systems like Feelex \cite{iwata2001feelex}, Lumen \cite{poupyrev2004lumen}, and inForm \cite{follmer2013inform} have demonstrated how digital data can be rendered tangibly, allowing users to feel, touch, and manipulate shape changes through hand and body engagement \cite{nakagaki2016materiable, leithinger2015shape, taher_2015_exploring}. For “force” feedback, inFORCE \cite{nakagaki2019inforce} introduced a bi-directional “force” display, enabling the detection and exertion of variable force patterns to simulate different material properties. Recent innovations have modularized these tabletop displays to create encounter-type haptics for hand or body exploration \cite{siu2018shapeshift, suzuki2019shapebots, suzuki2021hapticbots, suzuki2020lifttiles}. Miniaturized pin displays have also been developed for fingertip tactile recognition \cite{kammermeier2000dynamic, kim2009small, grigorii2022spatial, sarakoglou2012high}. In addition to grounded platforms, researchers have explored mechanical actuation methods with micro servo or miniaturized linear actuators to create handheld \cite{jang2016edge, yoshida2020pocopo, benko2016normaltouch}, and wearable haptic shape displays \cite{sarakoglou2005portable, huang2017retroshape}. \revise{Pneumatic actuator arrays have also been integrated into haptic sleeves \cite{Pasquier2023knithaptic} and gloves\cite{HaptX}.} Advancements in electro-actuated materials \revise{\cite{leithinger2023eam, acome2018hydraulically, shultz2023flat}} \revise{have significantly improved miniaturization and silent actuation. These inventions not only enable more compact tabletop shape displays \cite{johnson2023multifunctional, rauf2023electroadhesive} but also greatly enhance wearable haptics.} Fluid Reality \cite{shen2023fluidreality} introduces a self-contained haptic glove that achieves high-resolution shape changes on the fingerpad using electroosmotic pump arrays \cite{shultz2023flat}, \revise{while FlexEOP further explores the more flexible method \cite{Yu2024FlexEOP}.} In full-body haptics, the HAXEL system allows for thin, flexible haptic pin displays that can be attached to and perceived on different body parts \cite{leroy2020multimode, leroy2023hydraulically}.



Despite the promising potential of advanced pin displays for everyday haptics, their technical complexity hinders rapid prototyping. From a design perspective, our work aims to prioritize the sensory experience, enabling designers to explore what diverse and nuanced touch patterns can be rendered through pin-based shape displays and how the touch could engage with the human body. Therefore, a more accessible shape display specialized for exploratory design is necessary. Researchers have sought to make pin displays more approachable. For example, ShapeClip introduced light-responsive actuation modules that can transform any computer screen into a shape display \cite{hardy2015shapeclip}. While it significantly enhances the rapid prototyping of shape-changing interfaces, it may not be ideal for bodily haptics. Using fewer electronic components, the MorphMatrix \cite{dai2024morphmatric} toolkit allows 64 pins to be driven by 2 motors using a camshaft and lever mechanism, lowering the entry barrier for shape-changing design. However, each time, a unique camshaft must be fabricated. MagneShape \cite{yasu2022magneshape} and MagneSwift \cite{yasu2024magneswift} introduced passive magnetic pin displays that render animated shapes actuated by moving magnetic sheets, enabling rapid reprogramming and real-time manual sketching. However, these systems can only exert limited normal force while being gravity-dependent, restricting their ability to test dynamic force feedback across different body parts. Shape-Kit addresses these limitations with its long, flexible shape and force transducer, enabling an analog shape display that can manually render dynamic touch, allowing for free-form exploration across the body.


%% file: 3_design_of_shape-kit.tex

\section{Design of Shape-Kit}
Shape-Kit is a design toolkit to embody the “crafting haptics” design metaphor, facilitating the exploration of on-body expressive haptics. Based on the literature review in Section 2, we selected an analog, manually actuated pin-based shape display format. After building an early prototype and conducting first-person \revise{testing} (see Appendix \ref{section:early_prototype}), we iterated and refined the design with three goals: intuitive use, perceptible pin sensations across the body, and support approachable and comfortable collaborative haptic exploration. Additionally, we aimed to develop an ad-hoc tracking method to digitize the crafted patterns. In this section, we present the final design of the Shape-Kit toolkit, incorporating both analog and digital systems. \revise{We open source the toolkit by sharing the hardware components, 3D printing models, and software package\footnote{https://www.ranzhourobot.com/shapekit}.}

\begin{table*} [ht]
    \centering
\caption{Pin scales feature for displays in two Shape-Kit modules. Target areas are based on point localization threshold \cite{lederman2009haptic}}
\label{table_pinscale}
    \begin{tabular}{p{25pt}p{75pt}>{\raggedright\arraybackslash}p{55pt}p{135pt}p{160pt}} 
    \toprule
         \textbf{Module}&  \textbf{Pin Display Scale}&  \textbf{Target Area} &  \textbf{Potential Crafting Method}& \textbf{Spring Scale}\\ \midrule
         \textbf{M1}&  Small Display (S)

5x5mm&  Fingers, hallux, cheek

&  Use 1-2 fingers to actuate the entire display, creating uniform shapes amplified on the Large display& Coil diameter: 4mm; Free length: 15mm

Wire diameter: 0.3mm; \revise{Revolutions: 12revs

Spring constant: 0.1N/mm}\\ \hline 
        \textbf{ M2}&  Medium Display (M)

10x10mm&  Foot sole, calf, belly, forehead, forearm, palm&  Actuate 2-3 pins with one finger for controlled patterns, or use the palm to create uniform shapes amplified on Large display& Coil diameter: 4mm; Free length: 15mm

Wire diameter: 0.3mm; \revise{Revolutions: 12revs

Spring constant: 0.1N/mm}\\ \hline 
         \textbf{M1}\textbf{\&2}&  Large Display (L)

15x15mm&  Back, thigh, breast, upper arm&  Push individual pin or multiple pins with fingers or palm to create detailed patterns that can be minified to smaller scales& Coil diameter: 4mm; Free length: 8mm

Wire diameter: 0.3mm; \revise{Revolutions: 5revs

Spring constant: 0.3N/mm}\\
\bottomrule
    \end{tabular}
\end{table*}


\subsection{Shape-Kit Analog System}
Based on insights from early prototypes (Appendix \ref{section:early_prototype}), we developed Shape-Kit (Fig. \ref{fig:analog_system}), featuring two passive shape displays connected by flexible Bowden cables. Specifically, we used 914mm Gold-N-Rods cables\revise{\footnote{Sullivan Cable Type .032 Gold-N-Rods Rods 36 inch}}, which consist of a stainless steel multistrand internal cable that moves smoothly within a nylon tube. Bowden cables transmit linear motion and force to a spatially distant and extended point from the actuation side, which has been employed in tabletop shape displays \cite{follmer2013inform, nakagaki2020transdock, kammermeier2000dynamic}. We leveraged its flexible feature to explore the hand-held potential. This design allows the touch perception area on the body to be distanced from the crafting hands, enabling more flexible body positioning for solo use and potentially reducing social discomfort during collaborative prototyping.

The shape displays are spring-back on both sides, enabling bi-directional crafting and rendering. We used stainless steel springs instead of the custom ones in Appendix \ref{section:early_prototype} to accommodate the increased size and weight of the pins that required more resilient springs. \revise{The chosen springs are still soft and spongy enough to allow flexible actuation (Table \ref{table_pinscale}). }To provoke exploratory outcomes, we intentionally introduced ambiguity \cite{zhou2023tactorbots, huang2010mobile} by incorporating mismatched scales between the 5x5 pin arrays on each side. The cubic-shaped pins, with minimal spacing, support fluid actuation and continuous shape rendering. Pin scales were determined based on point localization thresholds \cite{lederman2009haptic} while also accommodating various crafting strategies (Table \ref{table_pinscale}). Shape-Kit includes two modules, both with large displays, paired with either small (M1) or medium displays (M2). \revise{They are lightweight and portable (M1: 290g, M2: 340g).} By simply pressing or rubbing with hands on one side, Shape-Kit can amplify (or minify) the crafted touch signals into pin-based haptic patterns transduced through the long cables. The output display can be moved around to test on different body parts or on another person’s body. 

Shape-Kit is modular and reconfigurable (Fig.\ref{fig:analog_system}), fabricated by FDM 3D printing. The small display can transform into a medium size by magnetically snapping pin caps, while the housings for the displays are also swappable. The large display can swap frames (e.g., a frame attached with velcro was used in the user study), with future iterations potentially adopting the same modular approach as the smaller displays. The default pin tips are flat, but other shapes of tips can be attached via embedded magnets or flexible fabric covers. This reconfigurable feature enhances Shape-Kit’s versatility as a haptic design tool.

\subsection{Shape-Kit Digital System}
\subsubsection{\revise{Tracking System}}
\begin{figure*}[h]
  \includegraphics[width=\textwidth]{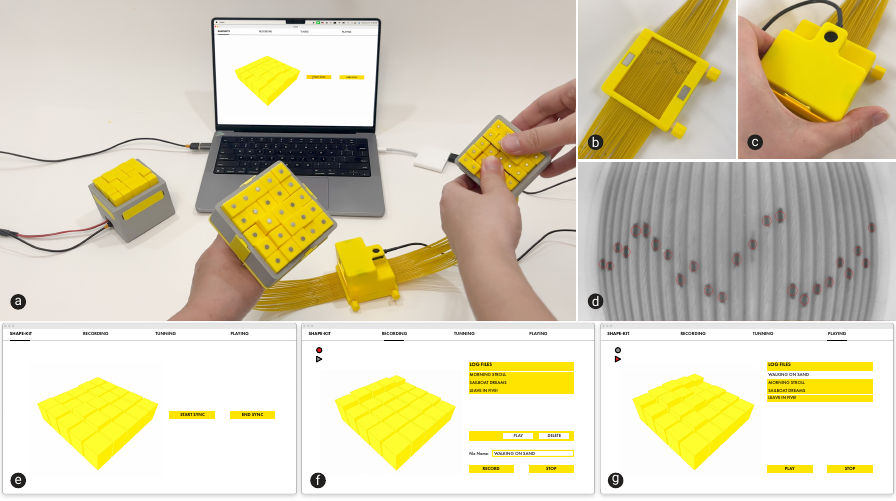}
    \caption{Tracking System and Shape-Kit GUI. (a) Synchronization on GUI (b) Window module (c) Tracking module (d) Computer vision tracking view (e) Tracking Synchronization Page (f) Pattern Recording Page (g) Pattern Playback Page}
  \Description{(a) a photo shows how the tracked pattern can be synchronized onto the GUI with 3D simulation. In the photo, there is a laptop running the Shape-Kit GUI, and there is a set of ShapE-Kit being actuated by a person with hands. (b) a close-up of the window module. (c) the tracking module is attached to the window module with embedded magnets (d) computer vision tracking view, showing how those marks on the cable can be captured by the tracking program. three screenshots for the Shape-Kit GUI: (e) tracking synchronization page: where there is a simulation of the shape display together with two buttons (f) pattern recording page: where there is a simulation with a list of documented touch patterns, and the type in box and buttons (g) pattern playback page. Where there is a simulation with a list of documented touch and play or stop buttons }
  \label{fig:tracking_system}
\end{figure*}

The Shape-Kit tracking system was inspired by Skinflow \cite{soter2019skinflow}, which used liquid transmission and optical sensors to measure pressure. Instead of dyed liquid, we leveraged Bowden cable’s mechanism, where the internal cable moves within a semi-transparent tube. 
As the pin’s actuation motion is transmitted through the cable, we applied dark markers (spaced 100mm apart) on the internal cable (Fig. \ref{fig:analog_system} \& \ref{fig:tracking_system}b\revise{)} to make the cable displacement more visible, enhancing precise computer vision tracking. The Bowden cables for each pin are meticulously arranged and secured in a window module, 
which allows clear detection of each pin’s movement to map them into a data array. 
The window module can be moved along the cable and tightened using side knobs.

Designers can document crafted sensations by magnetically mounting the tracking module (Fig. \ref{fig:tracking_system}c). After testing several optical sensors, we found a wide-angle webcam ideal for short-distance, low-latency tracking\footnote{HD 1080P USB CCTV Mini Security Camera (2.8 mm lens, 120-degree angle)}. We also integrated a portable camera light into the module to ensure stable lighting. \revise{The weight of the tracking module, including the webcam and light, is 90g.} To initiate tracking, the designer first turns on the light, connects the camera to a PC via USB and runs our custom Python script that opens a camera tracking view for calibration and detection. \revise{The vision system operates continuously to take image frames at a 30-Hz frame rate, detect the coordinates of each marker in the frame, and map them to a data array. By calibrating and calculating the relative changes in the marker coordinates, the vision system accurately tracks each pin’s movement.} They can then be proceeded by Shape-Kit GUI.

\subsubsection{Graphic User Interface}
Shape-Kit GUI used in our design study (Fig. \ref{fig:tracking_system}) was built with Processing and running on a PC, featuring a digital simulation of the shape display. The GUI accesses the Python script for tracking, allowing designers to synchronize the digital simulation with the touch patterns crafted on analog Shape-Kit in real-time. When pressing the “Start Sync” button, the software automatically calibrates the first frame as the baseline. Subsequent movements are tracked as displacement differences from this baseline and are mapped to each pin’s height, which is displayed in the 3D simulation. 

Designers can also digitally record and playback shape patterns, with additional controls for tuning general pin height and motion speed. We also developed a web-app version of the GUI for enhanced usability and stability. Both software applications’ packages are open-sourced.


\subsubsection{Programmable Shape Display} 

\begin{figure}[h]
  \includegraphics[width=0.48\textwidth]{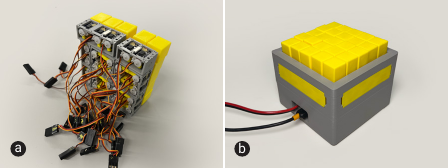}
  \caption{Servo-driven programmable shape display.}
  \Description{(a) the internal view of the computational shape display, which shows those modular units driven by micro linear servo. (b) a photo for the computational shape display in use}
  \label{fig:low_fi_display}
\end{figure}

To fully demonstrate the “crafting haptics” design metaphor, we developed a programmable shape display (Fig. \ref{fig:low_fi_display}) matching the scale of the large analog Shape-Kit display (15×15mm) and weighing 330g. Each pin is actuated by a micro plastic-gear linear servo \footnote{AGFRC 9mm coreless digital linear servo C1.5CLS-Pro} and arranged in modular units for easy reconfiguration and repairs. The display is entirely 3D-printed and controlled by an Arduino Nano with two Adafruit PCA9685 16-channel servo drivers. When connected to the Shape-Kit GUI, it can playback recorded touch patterns in tangible form.

We acknowledged the display’s technical limitations (e.g., limited torque of 240gf@6.0V with less robust plastic gears) but chose not to incorporate more advanced technology for two reasons: first, our focus was on the crafting method itself; second, building a custom shape display remains technically complex. Even with all components in place, assembling, programming, and debugging this “simple” programmable display took the first author three weeks, whereas the analog Shape-Kit was assembled in a single day and functioned immediately.

In the following study, we encouraged designers to focus on using the analog Shape-Kit, with outcomes documented via the tracking module and GUI. Designers tested the servo-driven display only during the system introduction to understand the concept and after completing their explorations to evaluate tangible playback. Despite its technical limitations, the display effectively demonstrates the crafting, recording, and playback process within the design system and has the potential to inspire future haptic exploration using similar approaches with more advanced technologies, such as the wearable shape displays discussed in Section \ref{sec:pin-based}.


%% file: 4_haptic_design_study.tex
\begin{figure*}[h]
  \includegraphics[width=\textwidth]{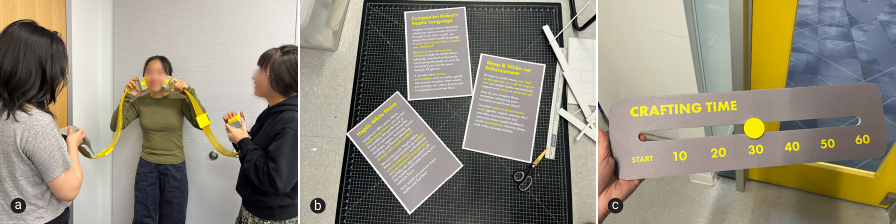}
  \caption{Pilot study and new study materials. (a) Collaborative prototyping in pilot 2 (b) Tangible posters for the study prompts (c) Tangible time indicator}
  \Description{(a) a photo from the pilot study 2, where 2 participants were crafting a Shape-Kit respectively and the 3rd participants put both output ends onto her cheek to test out the sensations. (b) a photo shows the making process for the posters that present our inspirational topics and prompts. (c) a close-up of our tangible time indicator, which is a timeline with a tangible slider}
  \label{fig:pilot_study}
\end{figure*}

\section{Haptic Design Study}
We conducted a design study to characterize how crafting with Shape-Kit facilitates expressive haptic exploration. The study was structured as small group sessions (2-3 participants) to ensure a collaborative, fully engaged, yet comfortable design experience. To inspire exploration, we provided 3 design topics with prompts (see Section \ref{section: study_procedure} \& Appendix \ref{section:topics with prompts}). Our study aimed to capture participants’ design experience through bodily engagement and discussion. We were particularly interested in how they developed various touch design ideas, crafted diverse sensations with Shape-Kit, and explored touch across different body parts.



\subsection{Study Preparation and Setup}

\begin{figure*}[ht]
  \includegraphics[width=\textwidth]{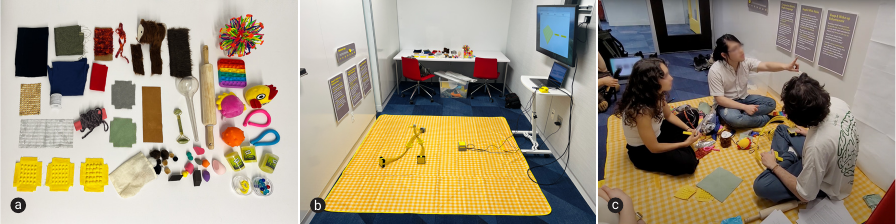}
  \caption{Study preparation and setup. (a) Provided props and materials (b) The space setup (c) Sample study scene from T2}
  \Description{(a) a photo shows all the props and materials used in our study. Which includes various fabric swatches, makeup spongy, slime, and some fidget toys (b) a photo of the study setting, which has a picnic mat on the floor of the conference room. (c) a photo of participants engaging with Shape-Kit in the study. The three participants in team 2 are sitting on the mat and discussing on the design topics while two of them having Shape-Kits in hands for prototyping}
  \label{fig:study setting}
\end{figure*}

We first conducted two rounds of pilot studies (Appendix \ref{section:pilot_study}, Fig. \ref{fig:pilot_study}a) to refine the design and structure of our exploration sessions. Building on insights from Pilot Two, we prepared additional props and materials (Fig. \ref{fig:study setting}a), alternative handles (Fig. \ref{fig:teaser}, Fig. \ref{fig:analog_system}), and continuously invited participants to play with slime at the start of each session to sensitize their hands \cite{hook2018designing}. The pilots also emphasized the need for a more flexible environment. Inspired by soma design practices \cite{hook2018designing, haynes2023meaningful}, we transformed a small conference room into a cozy space with a large picnic mat, allowing participants to choose comfortable postures during the design process. Researchers also sat on the floor to foster better communication (Fig. \ref{fig:study setting}b\&c). Based on pilot feedback, we printed topic prompts (Appendix \ref{section:topics with prompts}) on foam board posters for better readability and designed a tangible time indicator (Fig. \ref{fig:pilot_study}b\&c). They both have embedded magnets for easy repositioning on the whiteboard. To minimize disruptions from verbal time reminders, a researcher silently moved the slider on the time indicator every 10 minutes.

Two analog Shape-Kit modules were provided for the study. While the displays were swappable, participants did not reconfigure them due to time constraints, focusing instead on felt experiences and the design process. A laptop running the Shape-Kit tracking system and GUI was connected to a large monitor, with the tracking module and programmable display connected via USB. Each session involved three researchers: the first author acted as the primary facilitator, guiding and engaging with participants throughout the process and leading the interviews, while the other two managed space setup, technical moderation, time management, and video documentation. We used a fixed camera to record the entire session and a handheld camera for close-up highlights.


\subsection{Study Procedure}
\label{section: study_procedure}
Our study was approved with an exemption status by the University of Chicago’s institutional review board due to minimal risk. The study lasted 90 minutes and was video-recorded for analysis. The procedure was as follows:

\textbf{(1) Introduction:} Upon arrival, participants sat on the mat. After reviewing and signing the consent form, they were asked to use hand sanitizer. Then, each participant was given a can of slime to play with, which aimed at sensitizing their hands. While playing, they were asked to introduce themselves, focusing on their expertise and background in design and art. The facilitator then briefly introduced the Shape-Kit toolkit, the “crafting haptics” metaphor, and the study goal and plan. 

\textbf{(2) Shape-Kit Toolkit Warming Up:} Two Shape-Kit modules were placed on the mat for free exploration. As participants interacted with the tool for the first time, the facilitator provided prompts to encourage them to try out diverse crafting methods and body locations (see Appendix \ref{section:prompts in warming up}). The facilitator then guided them through the tracking and recording process using a test pattern they crafted. To demonstrate the full-cycle system, the recorded pattern was played back on the programmable shape display, allowing participants to experience it. However, throughout the study, they were encouraged to focus on crafting with the analog Shape-Kit.

\textbf{(3) Design Exploration:} Participants were presented with three inspirational topics with prompts for creating expressive haptics: \textbf{\textit{Sleep \& Wake-up Enhancement (SLE)}}, \textbf{\textit{Haptic White Noise (NOI)}}, and \textbf{\textit{Companion Robot’s Haptic Language (BOT)}}. Detailed prompts can be found in Appendix \ref{section:topics with prompts}. These topics were derived from a prior co-speculative exploration with domain experts (e.g., designers, researchers in haptics, human-robot interaction, and health) involving our earlier haptic design tool \cite{zhou2023tactorbots}. The prompts served as inspiration rather than prescriptive tasks, giving the team flexibility in how many topics they chose to explore. Each team had 60 minutes to ideate and prototype expressive haptic interactions, assign names to each pattern, and record their outcomes.

\textbf{(4) Post-study Evaluation and Interview:} 
Participants reviewed the digital simulation of their recorded touch patterns and tested the computational playback. Due to the scale of the programmable display, only patterns designed with the large display were played back. Participants would attach the same materials or fabrics to the programmable display as they had during crafting. They then completed a questionnaire to evaluate their Shape-Kit design experiences (see Appendix \ref{section:questionnaire}), followed by a semi-structured interview to provide qualitative feedback \ref{section:interview}).

\begin{table*}[]
    \centering
    \caption{Participant information in the haptic design study}
    \label{tab:participant_information}
    \begin{tabular}{p{17pt}p{60pt}p{35pt}p{50pt}p{75pt}p{208pt}}
\toprule

 \textbf{Team}& \textbf{Name}& \textbf{Pronoun}& \textbf{Relationship}& \textbf{Expertise}&\textbf{Areas of Interest}\\
 
\midrule
  \multirow{2}{*}{\textbf{T1}} &  Yifan Zou&  She/her&  Friend&  Artist, Maker& Drawing, multi-media installations, textile\\
 \cline{2-6}
 &  Yaochu Bi&  He/him&  Friend&  Designer, Artist& Multimedia and spatial design\\
\hline
         \multirow{3}{*}{\textbf{T2}} &  Leo Frankel&  He/him&  Friend of Mel&  Designer& Graphic design, typography, projection mapping\\
         \cline{2-6}
         &  Mel Damasceno&  She/her&  Friend of Leo&  Designer& Graphic design, typography, book and poster design\\
         \cline{2-6}
         &  Jamie Shiao&  He/him&  Stranger&  Artist& 2D art, game design, clay sculptuing, Blender for VR\\
         \hline
         \multirow{3}{*}{\textbf{T3}}&  Jasmine Lu&  She/her&  Stranger&  Researcher, Designer& Experienced in haptics, bio design\\
         \cline{2-6}
         &  Zoey Zhang&  She/her&  Stranger&  Designer& Background in art history\\
         \cline{2-6}
         &  Duruo Li&  She/her&  Stranger&  Artist& Illustration, poetry, oil painting, watercolor\\
         \hline
         \multirow{2}{*}{\textbf{T4}}&  Vivian Auduong&  She/her&  Friend&  Artist, Musician& Art, music, graphical simulation. VR/AR\\
         \cline{2-6}
         &  Ali Almiskeen&  He/him&  Friend&  Design Engineer& Industrial design, 3D printing, everyday product design\\
                 \hline
 \multirow{2}{*}{\textbf{T5}}& Aashina Songh& She/her& Couple& Designer, Artistr&Architecture, emotive active spaces,  phy-gital games\\
 \cline{2-6}
 & Arielle Schnur& She/her& Couple& Artist&Art history, performances, comedy, fiber and paper craft\\
         \hline
 \multirow{2}{*}{\textbf{T6}}& Emilie Faracci& She/her& Friend& Artist, Designer&Media art, photography, visual design, tangible interaction\\
 \cline{2-6}
 & Marcelle Brooks& She/her& Friend& Artist&Crochet, knitting, fiber art with tech\\
\bottomrule
    \end{tabular}
\end{table*}


\subsection{Participants}
We conducted six small-group study sessions with a total of 14 participants, who self-identified as female (10), male (4), aged between 18 to 35 years old (M = 23.36, SD = 4.67). \revise{Participants were} recruited through email lists from the Media Arts and Design Program at the University of Chicago and the School of the Art Institute of Chicago. Table \ref{tab:participant_information} details the participants’ information and backgrounds. All participants consented to video recording and released the names associated with their creations. As a design study, we aim to credit the designers for their original work and unique perspectives. Given that the study focused on touch design, we aimed to create a safe and comfortable space by explicitly considering participants’ preferences when forming teams. Thus, participants could sign up with friends or in same-gender teams. Most sessions included participants who attended in pairs, either as friends or, in Team 5, as a couple. Detailed relationships are listed in Table \ref{tab:participant_information}.

\subsection{Data Collection and Analysis}
We documented the study data with video recording, tracked touch patterns, and the questionnaire results. We transcribed the videos and compiled the close-up highlights. Since we were investigating the crafting of on-body haptics, we used hand sketching to document participants’ body placement exploration and crafting methods when reviewing the videos. We wrote thick descriptions of each team’s design process \cite{geertz2008thick}, color-coding the data to distribute behaviors, quotes, and reflections, helping us better understand and interpret the rich dataset. Using a grounded theory approach \cite{charmaz2006constructing}, we analyzed the data to identify key themes that emerged from participants’ sensorial exploration and design experiences in the study, which are presented in Section \ref{sec:results}.

%% file: 5_results_and_findings.tex

\section{Results and Findings}
\label{sec:results}
\subsection{Design Outcomes}
The participants generated 22 touch patterns, with each team designing 3 to 5. Among them, 9 were created for SLE, 9 for NOI, 3 for BOT, and 1 without a specified topic. Participants found the SLE and NOI topics particularly relatable, as they could easily envision them in their daily lives, effectively stimulating expressive haptic ideation. Many of them expressed excitement about the potential for their designed touch experiences to become an integral part of their routines, showing the willingness and desire to embrace this interactive medium. 

When crafting touch patterns with Shape-Kit, participants typically considered four key design elements: input/output display scale, output texture, body location, and the pin movement patterns, which were shaped by their actuation methods and crafting behaviors. Table \ref{tab:touch outcomes} presents the design outcomes from each team, highlighting their exploration of body placements and design features. Two patterns from each team were selected for detailed analysis of those design elements, chosen for their representation of each team’s typical design approach \footnote{Reference of human sketch design: https://www.dimensions.com/}. 

\indent
\begin{table*}
\caption{The haptic design exploration outcomes}
    \centering
    \begin{tabular}{c}
         \includegraphics[width=0.96\textwidth]{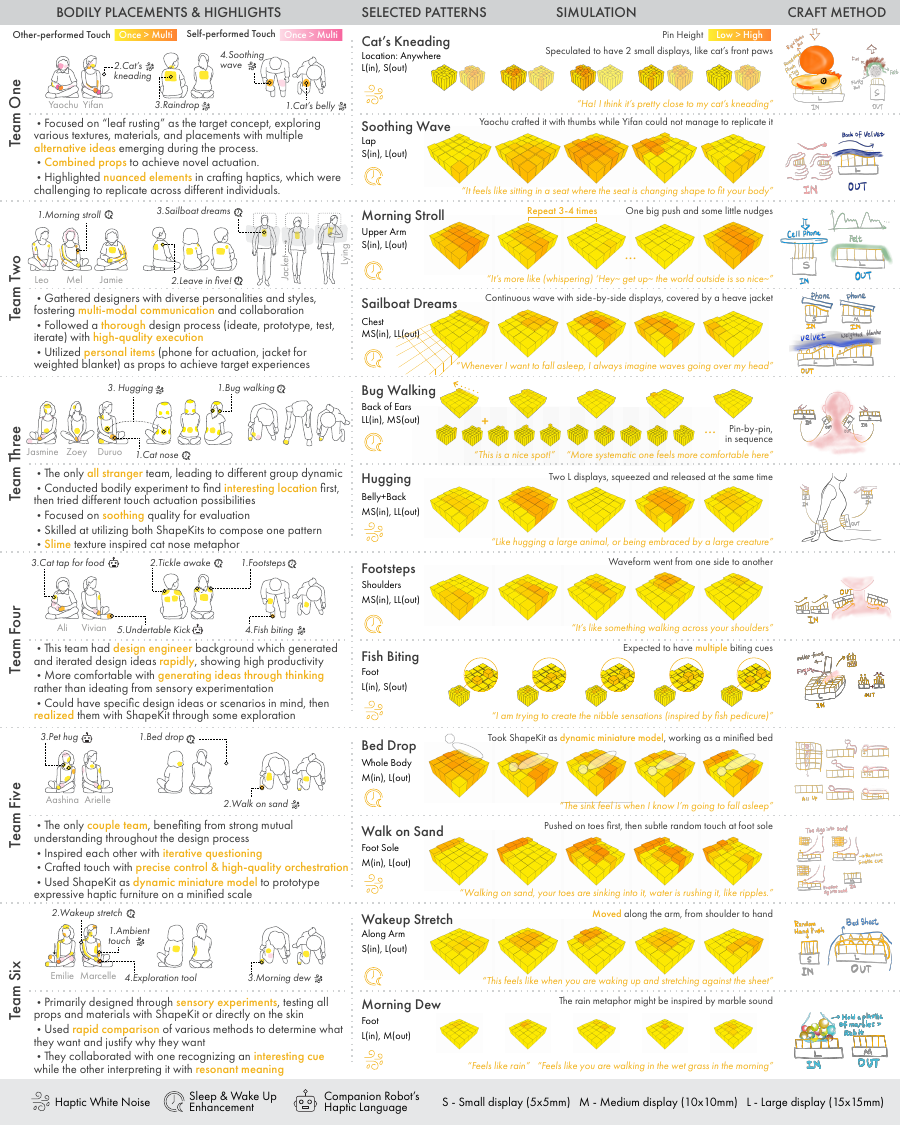}
    \end{tabular}
    \Description{This is a graphical table that presents the study results visually. The first column included the team number; the second column included body maps that show where each participant in each team explored touch on their body, and also highlights for each team; the third column shows two selected touch patterns from each team, included their name, target body placement, the scale of display used for in/output, the topic it was designed for, the right side including some key frames from the tracked pattern simulation; the fourth column included the hand sketches that show how participants crafted each touch pattern. 
Team One: 
•Focused on “leaf rusting” as the target concept, exploring various textures, materials, and placements with multiple alternative ideas emerging during the process.
•Combined props to achieve novel actuation.
•Highlighted nuanced elements in crafting haptics, which were challenging to replicate across different individuals.
Selected patterns: Cat’s kneading and Soothing wave
Team Two: 
•Gathered designers with diverse personalities and styles, fostering multi-modal communication and collaboration 
•Followed a thorough design process (ideate, prototype, test, iterate) with high-quality execution
•Utilized personal items (phone for actuation, jacket for weighted blanket) as props to achieve target experiences
Selected patterns: Morning stroll and Sailboat dreams
Team Three:
•The only all stranger team, leading to different group dynamic
•Conducted bodily experiment to find interesting location first, then tried different touch actuation possibilities
•Focused on soothing quality for evaluation
•Skilled at utilizing both ShapeKits to compose one pattern  
•Slime texture inspired cat nose metaphor
Selected patterns: Bug walking and Hugging
Team Four:
•This team had design engineer background which generated and iterated design ideas rapidly, showing high productivity
•More comfortable with generating ideas through thinking rather than ideating from sensory experimentation
•Could have specific design ideas or scenarios in mind, then realized them with ShapeKit through some exploration
Selected patterns: Footsteps and Fish biting
Team Five:
•The only couple team, benefiting from strong mutual understanding throughout the design process
•Inspired each other with iterative questioning
•Crafted touch with precise control & high-quality orchestration
•Used ShapeKit as dynamic Lego to prototype expressive haptic furniture on a minified scale
Selected patterns: Bed drop and Walk on sand
Team Six:
•Primarily designed through sensory experiments, testing all props and materials with ShapeKit or directly on the skin
•Used rapid comparison of various methods to determine what they want and justify why they want
•They collaborated with one recognizing an interesting cue while the other interpreting it with resonant meaning
Selected Patterns: Wakeup Stretch and Morning Dew
}
   
    \label{tab:touch outcomes}
\end{table*}

\subsubsection{Actuation Patterns}
The outcome touch patterns were digitized using our toolkit, with simulation screenshots in Table \ref{tab:touch outcomes} and videos in the supplementary materials. While the pin array has the potential to render various non-literal shapes, the patterns observed in our study generally fell into familiar categories such as rhythmic waves or stroking motions, sequential pin actuation (e.g., activating each pin one-by-one), uniform rising and falling shapes, locomotion of a form, and subtle random movements. 

Initially, we were curious whether more unconventional pin movements would emerge due to the organic nature of hand-craft prototyping. The pin-array format, for example, seemed promising for exploring unique haptic illusions, such as the “cutaneous rabbit \cite{geldard1972cutaneous},” to inspire creative touch stimulations. However, we found that achieving such exploration was challenging for designers new to haptics. We also realized that pursuing “shape novelty” should not be the primary goal, particularly when investigating “how to touch well” and designing for everyday expressive haptics. Instead, “conventional” touch patterns, likely favored for their inherent pleasantness, offer a solid foundation. 

One insight gained from our study is that subtle variations within familiar patterns can create distinct sensory experiences suited to different contexts. For example, the sensation of pressing can vary significantly depending on whether it’s crafted by a bare hand, a rigid shape, or a spongy prop, each delivering a unique force pattern and perception. A waveform on the thigh crafted by two thumbs side-by-side might feel like an interactive seat “\textit{attempting to shape itself to fit your body}” (T1, see \ref{result:bare_hand}), while the similar wave on the chest under a weighted blanket could evoke the soothing sensation of Sailboat Dreams lulling a person to sleep (T2, see \ref{result:props_as_tool}). By using a manual crafting tool like Shape-Kit, designers can fine-tune these patterns, exploring alternative contexts, placements, forces, textures, and other nuances to create an expressive touch experience whose felt quality matches the intended context and affords meaningful sensory engagement.

\subsubsection{Bodily Placements}
The body map in Table \ref{tab:touch outcomes} documents the body placements explored by each participant, highlighting the final placements for registered touch patterns, which included arms, shoulders, thighs, foot soles, back, belly, and more. We noticed NOI patterns were often designed for lower body parts, likely due to the prompt indicating a working scenario while the brain and hands were occupied. For the SLE topic, waking-up patterns were typically placed on the shoulder and arm, while sleep-related patterns covered larger areas. For example, T2 envisioned a shape-changing weighted blanket for the chest, and T5 imagined a bed that could “\textit{sink}” itself. BOT patterns did not show distinct preferences. 

In addition to the final placement, we documented body areas explored during ideation and iteration, involving self- and other-performed touch, depending on who crafted the touch. While most locations were explored using Shape-Kit, participants also used props, materials, or even their own hands to experiment with the sensory experience (e.g., Mel (T2) tried combing her hair for SLE), which were also captured on the body map. The data reveals that participants explored a wide range of body locations, extending beyond typical social areas \cite{suvilehto2015topography}, while no interpersonal touch was observed except with T5, who were a couple. 

\revise{From our observations and interviews, we identified features of Shape-Kit that potentially encouraged openness in exploring alternative body placements. The first is the far proxemic distance with touch transduction. The transducer allowed touch crafting from a distance, which helped respect personal boundaries \cite{urakami2023nonverbal}. For instance, Jasmine shared: “\textit{There was a moment when you (the display holder) came up to me and pressed it onto my belly. I was a bit freaked out. So, I do think the distance helped a lot.}” We observed that the touch receiver could occasionally hold the output display by themselves, especially for more intimate body areas (Fig. \ref{fig:interesting_placement}). Participants noted that holding the display against their body while perceiving the dynamic sensation did not cause significant distraction, possibly due to the rapid adaptation of haptic perception during steady holding \cite{roudaut2012touch}. 

The second is the otherness of pin-filtered hand touch. Although the sensations created through Shape-Kit were primarily “hand-crafted,” participants described the perceived sensations as “\textit{inhuman}” (T2), “\textit{object or machine-like}” (T3), and “\textit{neutral}” (T3). Because the touch was “\textit{intermediated}” (T1, T3, T4), crafting touch through Shape-Kit “\textit{felt less like directly touching the person}” (T6). Instead, it felt like “\textit{trying to send a message to a person through touch or having the receiver be the audience or judge to try out the designed touch pieces}” (T1). This perspective aligns with the growing appreciation of otherness in the community \cite{zhou2023tactorbots, bewley2018blonut, hassenzahl2020otherware}, where robotic touch could move beyond anthropomorphic or zoomorphic paradigms, opening alternative design possibilities. Our findings reveal that hand-crafted touch, when transduced and filtered into a pin-based format, acquires a sense of otherness. These insights could inspire future expressive haptic designs for less commonly explored body areas or even more intimate regions, while ensuring the design process and resulting touch sensations feel “dignified” \cite{zheng2024towards}.
}


    
    


\begin{figure*}[ht]
  \includegraphics[width=\textwidth]{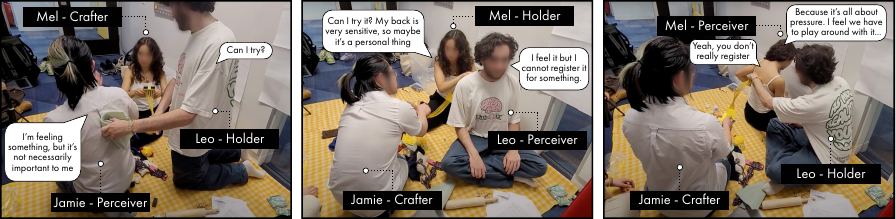}
  \caption{The role-switching process during T2’s sensorial exploration}
  \Description{photos and quotes to show the role-switching process in T2. Quotes: 
(a) Jack: ‘I’m feeling something, but it’s not necessarily important to me’
Liam: ‘Can I try?’
(b) Liam: ‘I feel it but I cannot register it for something.’
Mia: ‘Can I try it? My back is very sensitive, so maybe it’s a personal thing’
(c) Mia: ‘Yeah, you don’t really register’
Mia: ‘Because it’s all about pressure. I feel we have to play around with it…’}
  \label{fig:role_switching}
\end{figure*}


\subsection{Sensorial Exploration}

Participants employed a range of ideation methods, including mind mapping, metaphorical thinking, probing questioning, and drawing inspiration from multisensory experiences. Yet, it was the sensorial exploration that was central to their creative process. Shape-Kit’s embodied and tactile nature not only fostered this bodily exploration but also allowed ideas to emerge in ways that would be challenging in conventional brainstorming settings. Through somatic exploration, participants experimented with Shape-Kit’s capabilities, the tactile qualities of various props and materials, and the sensitivity and appropriateness of different body placements. As participants prototyped and tested their designs, they often noted that “\textit{what you imagine you might feel on your body and the real feeling is quite different}” (Duruo, T3). Participants also found this unpredictability to be one of the most appealing aspects of haptic design, resonating with how alternative perceptions can “foster more creative and reflective interpretations,” as proposed in prior research \cite{zhou2023tactorbots}. The collaborative aspect of Shape-Kit further enhanced this process, fostering touch iteration, interesting placement identification, and idea emergence.


\subsubsection{Collaborative Touch Iteration}
\label{sec:role_switching}

Some teams took frequent role-switching to ensure all members grasped the capabilities and limitations of Shape-Kit, leading to collaborative design decisions through iteration. T2 exemplified this approach (Fig. \ref{fig:role_switching}). When designing a gentle wakeup sensation for SLE, they began by testing on the back, envisioning the haptic system integrated into a bed. Mel crafted the medium display with wave patterns and sequential pin actuation, while Leo held the large display against Jamie’s back, using felt fabric as a mediator. However, Jamie remarked, “\textit{I’m feeling something, but it’s not necessarily important to me,}” prompting the team to iterate by switching the output side, changing textures, attaching tips, and adjusting the actuation method (e.g., actuate one or two pins with stronger force.) Despite these adjustments, the sensations on the back were still less impactful than expected. To ensure everyone was on the same page, they continued switching roles. Mel, the initial crafter, suggested: “\textit{Wait, can I try it? My back is very sensitive, so maybe it’s a personal thing.}” After testing it herself, she concluded, “\textit{Yeah, you don’t really register.}” Acknowledging the challenges in designing for back, she also noted, “\textit{because it’s all about pressure. We have to play around with it}” Leo then proposed trying the shoulder, which led to comparisons between sensations on the shoulder, back, and forearm. They found the forearm provided a more noticeable experience. 

As they refined the pattern, the team ruled out a rolling motion, finding it too soothing for waking up. Mel suggested, “\textit{I’m thinking about something that is very slow and gentle, kind of like…}” (she gently tapped her thigh 3 times). Leo and Jamie understood her idea and experimented with props, like a rigid ball and a mobile phone, to create a more uniform yet gentle tactile pattern. After confirming the sensation with everyone through role switching, they crafted a pattern described as “\textit{wakeupable}.” It featured a big push followed by little nudges. Mel captured the sentiment: “\textit{It’s more like (whispering) ‘Hey~ get up~ the world outside is so nice’}~” They named the design Morning Stroll.

\begin{figure*}[hbt!]
  \includegraphics[width=\textwidth]{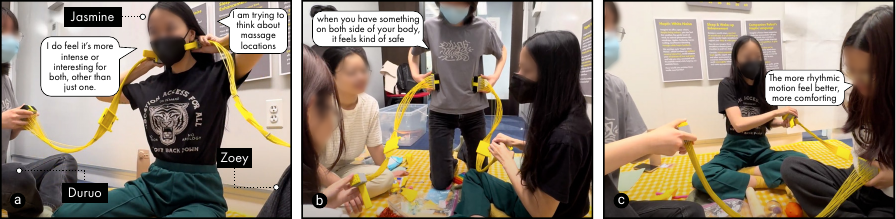}
  \caption{T3 prioritized using Shape-Kit to discover “interesting body placement,” particularly interested in using both Shape-Kits simultaneously on symmetric sides of the body. (a) Bug Walking, a soothing systematic sensation on the back of ears (b \& C) Placement exploration that led to the Hugging pattern}
  \Description{photos and quotes to show how T3 discover “interesting body placements. Quotes:
(a) Jade: ‘I am trying to think about massage locations.’
Jade: ‘I do feel it’s more intense or interesting for both, other than just one.’
(b) Dana: ‘when you have something on both side of your body, it feels kind of safe’
(c) Zoe: ‘The more rhythmic motion feel better, more comforting’
}
  \label{fig:interesting_placement}
\end{figure*}

\begin{figure*}[hbt!]
  \includegraphics[width=\textwidth]{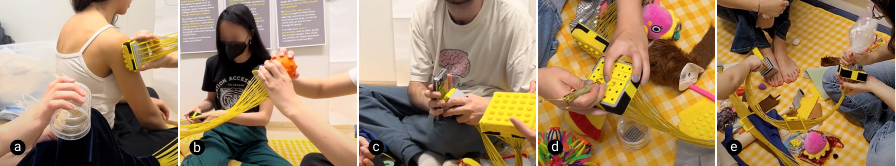}
  \caption{Participants used various props as tools to facilitate crafting. (a) Roll a cylinder (b) Roll a rigid ball (c) Press and tilt with a phone (d) Use the facial roller’s fork-shaped frame (e) Roll a plastic bag filled with marbles}
  \Description{photos show participants use various props as tools to facilitate crafting}
  \label{fig:props_as_tools}
\end{figure*}

\subsubsection{Placement Identification and Idea Emergence} 
\label{sec:placement}
Some teams relied heavily on sensorial exploration to spark their ideation process. For instance, Team 3 prioritized using Shape-Kit and props to discover “\textit{interesting body placements.}” Rather than analyzing logical locations for context, they freely explored random body parts, often with one team member actuating random patterns while another moved the display around the perceiver’s body. Once someone identified a special area, they invited other members to try it out by switching roles. They were also particularly intrigued by using both Shape-Kits simultaneously on symmetric sides of the body, exploring various touch patterns for their soothing effects (Fig. \ref{fig:interesting_placement}). \revise{In such cases, the touch receiver used both hands to hold the output displays and attach them to their body, while the other two designers worked as crafters, each actuating a Shape-Kit module.}

Team 6, a pair of two, embraced sensorial exploration by frequently switching roles with support from the study facilitator. There was a time when the facilitator crafted random patterns upon request. Meanwhile, Emilie moved the large output display, covered with two layers of tips beneath a bed sheet, across Marcelle’s body. She tested areas like the waist, lap, and shoulder. When the display touched her shoulder, Marcelle remarked, “\textit{I like that weirdly!}” When they switched roles, Emilie joked that it felt like her doctor moving a stethoscope across her back. They once experimented with holding the display and moving it along the dorsal of the arm (from shoulder to forearm) while maintaining random pin actuation. Emilie described it as “\textit{waking up and stretching against the sheet, which could work for waking up!}” Followed by some comparison testing, they registered it as Wakeup Stretch. They particularly emphasized the importance of keeping both locomotion and tactile stimulation, as adding subtle pushing enhanced the richness of the sensation. 

Similarly, Emilie noticed “\textit{interesting}” tactile cues when Marcelle rubbed a plastic bag filled with marbles onto the large display. Emilie, perceiving the sensation through the medium display on her palm and forehead, remarked, “\textit{I like these tactile things. They make me really comfortable. They are tiny little sensations with some unpredictability.}” Upon switching roles, Marcelle associated the feeling with “\textit{rain.}” When they moved the sensation to their feet, they found it felt like “\textit{walking in the wet grass in the morning,}” leading them to name the pattern Morning Dew.


    
    
    


\subsection{Crafting Methods}
While Shape-Kit was designed to facilitate haptic creation through simple hand touch, our study revealed the limitations of bare-hand actuation and underscored the value of using props to fully harness the hand’s dexterity. The pin display’s form factor only allows for the transmission of normal force, primarily limiting bare-hand actuation to pressing. However, by introducing tools and props, the hands can perform more diverse behaviors, such as rolling a ball or cylinder across the pins, rubbing a sponge into the display, or using fork-shaped tools to precisely actuate specific pins (Fig. \ref{fig:props_as_tools}). These tools allowed crafters to maintain full control over the rhythm, force, and angle of their actions, while the output textures further enriched the sensory experience. The integration of props and textures significantly expanded the design possibilities with Shape-Kit, enriching dynamic force patterns and enhancing the touch quality. Conversely, Shape-Kit translated the material properties of these props into nuanced dynamic features, with the pin-based display animating static textures and imbuing them with new meanings. In this way, Shape-Kit and the materials fostered novel haptic creation in a mutually reinforcing manner. In this section, we introduced participants’ typical crafting methods and those tacit insights. 

\begin{figure*}[ht]
  \includegraphics[width=\textwidth]{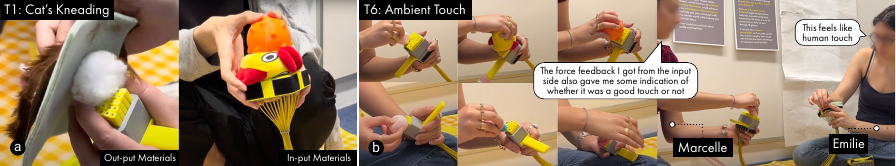}
  \caption{Participants utilized the material properties of props to enhance the tactile sensation. (a) T1’s crafting method for Cat Kneading (b) T6 compared a large range of props with various material properties to explore their felt qualities.}
  \Description{(a) the photos show cat’s kneading output end included a fluffy ball, a felt, and a layer of long fur fabric, the right side shows it was crafted with a round plush toy and a rigid ball (b) photos show how Maya tried crafting with various props for sensorial exploration}
  \label{fig:material_quality}
\end{figure*}

\begin{figure*}[ht]
  \includegraphics[width=\textwidth]{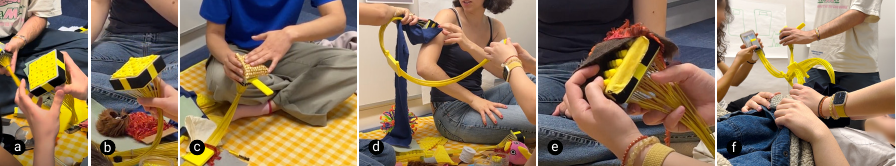}
  \caption{Delicate Textures Enhance Touch Quality. (a) A single tip for T2’s Leave in Five (b) Tips in “X” shape but found it alien by T6 (c) The metal mesh fabric inspired T5 with “sand” metaphor (d) Bed-sheet fabrics for T6’s Wakeup Stretch (e) T6 designed Animal Companion with layered textures \revise{(f) T2 incorporated jacket as “home prop” when prototyping Sailboat Dream}}
  \Description{photos show how participants employed various textures in enhancing the tactile quality}
  \label{fig:texture}
\end{figure*}

\subsubsection{Bare-hand Crafting with Nuances}
\label{result:bare_hand}

The hand’s dexterity supports a wide range of gestures, often explored in affective haptics \cite{hertenstein2009communication}. However, Shape-Kit’s form factor limits hand manipulations to normal force pressing. We had envisioned more dynamic motions inspired by musicians’ expressive techniques (e.g., vibrato, glissando) and bodyworkers’ (e.g., massage therapists) specialized touch techniques. However, these delicate controls were not observed in our study, suggesting that such techniques require special training or expertise.

Despite these constraints, bare-hand crafting revealed a wealth of subtle variations that, in some cases, were even challenging to replicate across individuals. These underscored the richness of touch perception that was uniquely captured through hand-crafted methods. At the same time, these observations highlighted the value of digital tracking, as such nuances would be challenging to document otherwise. 

For example, although many patterns used a waveform format, nuanced variations led to different perceptions and effects. In one case, Yaochu (T1) randomly actuated the small display with both thumbs. Yifan identified it as “\textit{a Soothing Wave!}” When she tried it on her thigh, she said, “\textit{It feels like you’re sitting in a seat where the seat is attempting to shape itself to fit your body.}” However, when they switched roles, Yifan observed Yaochu’s hand movements and attempted to replicate the wave sensation using various methods, from bare hands to rolling a plastic cylinder. Despite her efforts, she couldn’t recreate the “\textit{vivid and organic}” wave sensation she had previously experienced. The rolling cylinder, for instance, felt “\textit{too intentional and intense.}” \revise{They recognized the challenges of replicating hand-crafted nuances even within a small group. Therefore, though it could be distracting, Yaochu eventually tested the touch while crafting it himself. Yifan was assisting by holding the display against his thigh.}

Interestingly, when Yifan watched the real-time 3D simulation during the tracking process, she noted, “\textit{It actually looks better (clearer) on screen.}” She later explained that she had been more focused on watching Yaochu’s hand behaviors during the learning process; while the felt sensation differed, the shape variations were hard to distinguish by feeling alone. When visualized and enlarged in the 3D simulation, the pattern’s slow build-up and quick release motion became more apparent, likely due to Yaochu’s advanced muscle control when actuating with thumbs.


\subsubsection{Props as Tools for Enhanced Actuation}
\label{result:props_as_tool}

By leveraging their affordances, participants used props, much like sculpture tools in ceramics, to craft sensations that were difficult to achieve with hands alone. For instance, rolling a cylinder or a rigid ball to create more regular shape motions (Fig. \ref{fig:props_as_tools}a\&b), or using pens or sticks to enable precise pin actuation on smaller displays. To create holistic pin movement where all pins moved together, Leo (T2) looked for a large, rigid, flat surface and ultimately used his cell phone to actuate the Shape-Kit (fig. \ref{fig:props_as_tools}c). The cell phone, designed for handheld use, also facilitated fluid wave shapes by pressing down while slowly tilting, which they employed in crafting the Sailboat Dream. 

Participants also improvised custom tools to achieve specific shapes or sensations. For example, in designing the Fish Biting sensation, inspired by Vivian’s experiences in fish pedicure, T4 aimed to simulate a cohesive “\textit{nibble}” by pressing two pins separated by one and then pressing the middle pin. They realized that achieving a fast, synchronized action was challenging with fingers alone. They repurposed a facial roller’s fork-shaped frame to effectively actuate the two pins with the same motion (Fig. \ref{fig:props_as_tools}d). When trying it out on the foot sole, Vivian noted, “\textit{Now the sensation felt more like going together!}” Similarly, the Morning Dew pattern (T6) mentioned earlier was crafted with a plastic bag filled with marbles (Fig. \ref{fig:props_as_tools}e). These examples demonstrate how props can significantly enhance the range and subtlety of haptic interactions.

\subsubsection{Material Properties Enrich Force Pattern}

\begin{figure*}[ht]
  \includegraphics[width=\textwidth]{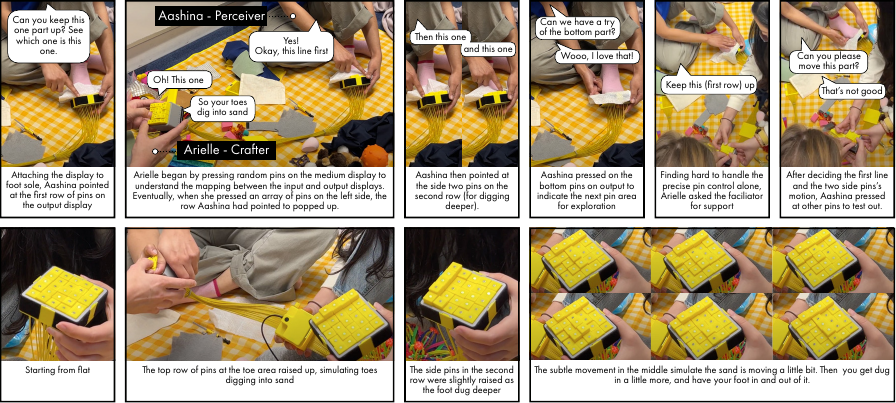}
  \caption{Design process of Walk on Sand: (1) Top row: Aashina provided real-time feedback verbally and through embodied cues to guide and coordinate precise pin control (2) Bottom row: the shape pattern of crafted touch}
  \Description{“a series of photos show how T5 crafted the walk on sand sensation with precise pin control, quotes and context:
(a) Attaching the display to foot sole, Ava pointed at the first row of pins on the output display. Ava: ‘Can you keep this one part up? See which one is this one.’
(b) Anya began by pressing random pins on the medium display to understand the mapping between the input and output displays. Eventually, when she pressed an array of pins on the left side, the row Ava had pointed to popped up.
Anya: ‘Oh! This one! Ava: ‘Yes! Okay this line first’
Anya: ‘ So your toes dig into sand’
(c) Ava then pointed at the side two pins on the second row (for digging deeper). 
Ava: ‘then this one, and this one’
(d) Ava then pressed on the bottom pins on output to indicate the next pin area for exploration.
Ava: ‘Can we have a try of the bottom part? Wooo, I love that!’
(e) Finding hard to handle the precise pin control alone, Anya asked the facilitator for support
Ava: ‘Keep this (first row) up’
(f) After deciding the first line and the two side pins' motion, Ava pressed at other pins to test out.
Ava: ‘Can you please move this part?’
Ava: ‘That’s not good’
The bottom row of photos show the shape pattern of their creation}
  \label{fig:sand}
\end{figure*}

\label{sec:property}
Participants creatively used the material properties of props (e.g., rigidness, sponginess, compliance, malleability) to enhance the touch quality rendered by Shape-Kit. Team 1’s crafting method for the Cat Kneading was notably interesting (Fig. \ref{fig:material_quality}a). The long fur fabric reminded Yifan of her cat. With a specific metaphor, they first combined a fluffy ball, felt, and fur fabric to create a soft, rounded output shape on the small display to mimic a cat’s paw. Meanwhile, Yaochu experimented with various props and approaches to actuate the large display. However, without the reference to an actual cat, his crafted sensation felt strong and serious, prompting Yifan to jokingly describe it as “\textit{feeling like a soldier is marching.}” After several rounds of role switching and refining, they ultimately used a combination of a rigid plastic ball with a round shape plush toy to achieve the “\textit{gentle and soft motion but in uniform shape}” that Yifan described as “\textit{feeling right}” and  “\textit{pretty close to my cat’s kneading.}” 

In another case, Marcelle (T6) used a makeup sponge on the medium display while Emilie perceived the sensation on her forearm, with a felt layer as a mediator (Fig. \ref{fig:material_quality}b). Emilie commented, “\textit{This feels like human touch.}” In contrast, when Marcelle used her bare hands to press the pins, Emilie noted, “\textit{This feels more like a toy.}” Interestingly, the sensation created by the sponge felt more human-like, likely due to its softer force, while hand pressing amplified the display shape too much. When asked if the touch felt uncanny \cite{mori2012uncanny}, Marcelle found it “\textit{comforting, surprisingly,}” possibly because it was behaviorally human-like instead of form-wise. 

As shown in Fig. \ref{fig:material_quality}b, T6 also experimented with various other props and display sizes for actuation. Marcelle, the crafter, noted that, beyond Emilie’s response on the perception, “\textit{The force feedback I got from the input side also gave me some indication of whether it was a good touch or not.}” This insight is particularly intriguing, as it highlights how playing the crafting role fully engages in sensorial exploration. As the crafter manually actuated the input display with hands or props, they could feel the push-back from the perceiver’s side, including friction from the skin, muscle tension, and the material properties of the props, such as sponginess and malleability. This tactile feedback allowed them to intuitively assess whether they had achieved a “\textit{good}” touch, drawing on their memory of previously crafted touches and their sense of what a pleasurable touch should feel like—not necessarily in shape, but in terms of force and rhythm.


\subsubsection{Delicate Textures Enhance Touch Quality} 

\begin{figure*}[ht]
  \includegraphics[width=\textwidth]{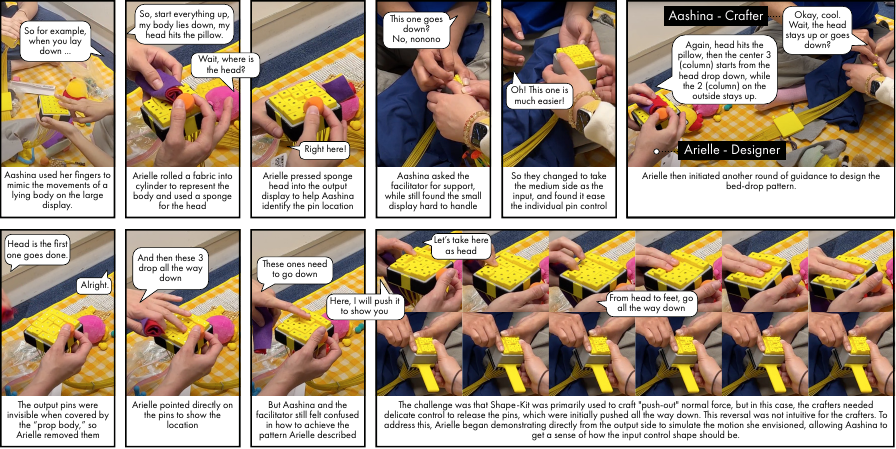}
  \caption{Design process of Bed Drop, where T5 used Shape-Kit as a dynamic miniature model to simulate a large-scale shape-changing interaction}
  \Description{a series of photos show how T5 designed Bed drop by using Shape-kit as dynamic miniature model, quotes and context:
(a) Ava used her fingers to mimic the movements of a lying body on the large display.
Ava: ‘So for example, when you lay down …’
(b) Anya rolled a fabric into cylinder to represent the body and used a sponge for the head
Anya: ‘So, start everything up, my body lies down, my head hits the pillow.’
Ava: ‘Wait, where is the head?’
(c) Anya pressed sponge head into the output display to help Ava identify the pin location
Anya: ‘Right here!’
(d) Ava asked the facilitator for support, while still found the small display hard to handle
Ava: ‘This one goes down?No, nonono’
(e) So they changed to take the medium side as the input, and found it ease the individual pin control
Ava: ‘Oh! This one is much easier!’
(f) Anya then initiated another round of guidance to design the bed-drop pattern.
Anya: ‘Again, head hits the pillow, then the center 3 (column) starts from the head drop down, while the 2 (column) on the outside stays up.’
Ava: ‘Okay, cool.Wait, the head stays up or goes down?’
(g) The output pins were invisible when covered by the “prop body,” so Anya removed them
Anya: ‘Head is the first one goes done.’
Ava: ‘Alright’
(h) Anya pointed directly on the pins to show the location
Anya: ‘And then these 3 drop all the way down’
(i) But Ava and the facilitator still felt confused in how to achieve the pattern Anya described
Anya: ‘These ones need to go down’
(j) The challenge was that Shape-Kit was primarily used to craft "push-out" normal force, but in this case, the crafters needed delicate control to release the pins, which were initially pushed all the way down. This reversal was not intuitive for the crafters. To address this, Anya began demonstrating directly from the output side to simulate the motion she envisioned, allowing Ava to get a sense of how the input control shape should be.
Anya: ‘Here, I will push it to show you’
Anya: ‘Let’s take here as head’
Anya: ‘From head to feet, go all the way down’
}
  \label{fig:bed_drop}
\end{figure*}

Participants explored a variety of textures to enhance the nuance of tactile sensations using alternative tips and fabric swatches (Fig. \ref{fig:analog_system}, Fig. \ref{fig:study setting}a), with different qualities (e.g., coarse/smooth, soft/rigid, spongy/solid, fluffy/plain, warm/cold). Different-shaped tips (round, sharp, and fluffy) were often applied to improve perceivability, with some applied magnetic tips selectively. Some participants attached a single tip to emphasize specific pin actuation, as seen in urgent wakeup designs (e.g., T2’s Leave in Five and T4’s Tickle Awake, see Fig. \ref{fig:texture}a). Others added tips on select pins to create geometric shapes, though these were sometimes perceived as alien due to their unconventional forms (Fig.\ref{fig:texture}b).

Fabric swatches were also a significant part of the exploration. Some participants drew inspiration from the textures (e.g., T5: Metal mesh fabric’s touch feeling inspired the “sand” metaphor, see Fig. \ref{fig:texture}c and Section \ref{sec:precise_control}), while others attached fabrics directly to the output display. Certain textures evoked specific metaphors: long fur was often associated with pets, bed-sheet fabrics related directly to the SLE topic (Fig. \ref{fig:texture}d), and felt or velvet were more likely to deliver soothing, gentle sensations. Participants also layered fabrics for more complex sensations. T6’s Animal Companion design, for example, combined three layers: sharp tips, a long-fur fabric on the reversed side for its fluffy properties, and tassel threads on top. They found it interesting when the pins only made contact when moving upward (Fig. \ref{fig:texture}f). Marcelle noted, “\textit{You kind of feel it, and especially the tassels, this feels nice.}” Emilie added, “\textit{The less touch area makes it more noticeable. When it’s the whole thing, you block it out, but when just a tiny bit brushes against you, it’s highlighted.}” 

However, the large, continuous fabric swatches often masked delicate pin movements. Unlike magnetic tips designed for individual pins, attaching a fabric swatch to the display would raise a continuous area with each pin actuation. Thus, for sensations where precise or delicate pin movements were crucial, participants often used the bare Shape-Kit or only with magnetic tips. 

\revise{Beyond the provided props and materials, participants even brought in “home props” to enhance the contextual quality of their designs. For instance, when ideating the Sailboat Dream for lulling sleep, T2 envisioned a haptic weighted blanket on the chest. To simulate this, Jamie lay fully down on the mat and grabbed his thick jacket to cover his chest (Fig. \ref{fig:texture}f). They prototyped the design by placing two large displays side-by-side to render a continuous wavy flow, layering the jacket on top of the dynamic pattern. The lying-down posture, combined with the jacket’s size and weight, greatly enhanced the scenario’s realism, resulting in a high-quality execution of the touch outcome.}

\subsection{Haptic Composition}
Most of the resulting touch patterns were “haptic phonemes” \cite{Enriquez2006phonemes}, cycling through repetitive sample phrases. Many of these patterns were designed for a single contact point, while others incorporated two symmetric (T3, see Fig. \ref{fig:interesting_placement}), or continuous contact points (Sailboat Dreams - T2, Footsteps - T4). Participants also used Shape-Kit to craft a haptic phoneme, envisioning that multiple instances could be rendered in a real implementation. For example, T1’s Cat Kneading pattern was intended to “\textit{have two of them press down in sequence,}” while T4’s Fish Biting was imagined as “\textit{multiple, probably a lot of fish.}” Some designs also involved locomotion across the body, such as T6’s Wakeup Stretch, which combined tactile stimulation with movement along the arm. T5 took a unique composition approach by focusing on narrative patterns and using Shape-Kit creatively as a dynamic miniature model to prototype whole-body touch experiences. We detail their design process below.




\subsubsection{Orchestration with Precise Control}
\label{sec:precise_control}
T5 took a unique approach by crafting a narrative pattern with precise pin control for high-quality orchestration. For the NOI topic, Aashina envisioned a scenario: “\textit{Yeah, walking… when you’re sitting and doing work on the computer, you may want to take off your socks and feel the grass and even feel the grass moving.}” As they experimented with materials, Arielle rubbed metal mesh fabric with her hands (Fig. \ref{fig:texture}c) and tried it on her foot sole, discovering, “\textit{This could be sand, like you’re putting your feet in the sand on the beach!}” 

This sparked them to “\textit{simulate the wet sand on the beach}” using Shape-Kit. They attached the large display to Aashina’s foot sole while Arielle manipulated the medium display. Aashina provided real-time feedback, both verbally and through embodied cues. She initially pointed to the desired pin locations on the output display. But when Arielle struggled to map the corresponding pins on her side, Aashina pressed the pins directly on the output display, utilizing Shape-Kit’s bi-directional feature to help Arielle locate the actuation areas. This also allowed them to precisely coordinate movement rhythm and pressure. Figure \ref{fig:sand} showed the coordinating process and the shape pattern. Through this meticulous process, they created a sensation where the top row of pins at the toe area raised up, simulating toes digging into the sand. The side pins in the second row were slightly raised as the foot dug deeper, with random, subtle movements added to the pins at the sole to mimic the texture of sand. When they switched roles, Arielle exclaimed, “\textit{I love this. I think that feels like the sand when you’re walking on it, and it feels uneven, like the ripples, and the water rushing in and changing it every time.}” The touch crafted by T5 stood out for its progression, closely matching the scenario and creating a dynamic, evolving sensory experience.


\begin{table*} 
    \centering
    \caption{Participants’ self-reported rating for evaluating Shape-Kit design experiences (M: Mean, SD: Standard Deviation)}
    \label{tab:self-report}
    \begin{tabular}{p{65pt}p{355pt}p{20pt}p{22pt}}
\toprule
 \textbf{System}& \textbf{ Questions}: Likert-scale rated from 1 = Strongly disagree to 5 = Strongly agree& \textbf{M}& \textbf{SD}\\
\midrule
  \multirow{3}{*}{\textbf{Analog}} &  I felt very confident using the system&  4.20&  0.40\\
 \cline{2-4}
 &  I thought designing haptic experiences with Shape-Kit was intuitive&  4.50&  0.33\\
 \cline{2-4}
         &  I believe Shape-Kit enables versatile haptic design&  4.50&  0.38\\
\hline
         {\textbf{GUI}} &  I believe the digital simulation on the Shape-Kit App can represent our design well&  4.30&  0.30\\
         \hline
         {\textbf{Programmable}} &  I believe the replay of the touch pattern can render our design well&  2.90&  0.62\\
\bottomrule
    \end{tabular}
\end{table*}


\subsubsection{Shape-Kit as Dynamic Miniature Model}
\label{sec:miniature_model}
Team 5 also introduced a novel approach to using Shape-Kit, treating it as a “dynamic miniature model” to prototype a complex, body-scale haptic experience. The design was inspired by Arielle’s personal experience: “\textit{For me, when I lay down in bed, and then you feel the sink… That ‘sink’ feeling is when I know I’m going to fall asleep, like that heaviness.}” The team discussed how to simulate this “\textit{sinking}” sensation, finding it challenging to convey through verbal or body language alone. Arielle suggested: “\textit{You know how there are weighted blankets that go on top of you? I wonder if there’s a way to ‘weight’ the bed. So you lay down, and then the bed drops, but not fully cocooning you.}” Aashina, who has an architecture background, attempted to use her hands to actuate the Shape-Kit to describe the movement in her understanding in a miniature form, “\textit{So, it’s like a little bit wavy pattern per se, or a depression, slowly.}” Inspired by this, they then decided to use Shape-Kit to create a minified model, with Aashina first using her fingers to mimic the lying down motion, while Arielle rolled a fabric into the cylinder to represent the body and used a sponge for the head. Fig. \ref{fig:bed_drop} showcases how Arielle guided the process. Since they aimed to simulate a “\textit{sinking}” motion, Aashina, the crafter, needed to start by pushing the input pins all the way down and then delicately release them, reversing the usual “push-out” motion Shape-Kit was designed for, making it less intuitive. To help, Arielle demonstrated using her two fingers to mimic the body and push down directly from the output side, allowing Aashina to sense how to control the input end.  

Together, they carefully adjusted the pins to simulate a bed that gradually sinks as the body lies down, starting from the head and slightly moving down to the feet, then dropping all the way down under the body area to create a “\textit{sinking}” sensation for lulling sleep. The use of Shape-Kit as a dynamic, scalable tool enabled the design to translate abstract conceptual ideas into a tangible, layered haptic experience, which also facilitated the idea communication within the team, demonstrating the potential of such tools in designing complex, large-scale touch and shape-change interactions.

\subsection{User Feedback on Crafting Haptics with Shape-Kit Toolkit}
Table \ref{tab:self-report} summarizes the results from the post-study questionnaire (Appendix \ref{section:questionnaire}), which included several Likert-scale questions assessing user experiences, with ratings ranging from 1 (strongly disagree) to 5 (strongly agree). The data indicates generally positive feedback on the usability and versatility of the analog Shape-Kit, with mean ratings of 4.20 or higher. Participants also rated the 3D simulation of their crafted patterns positively (M=4.3, SD=0.30). However, the tangible playback received lower ratings (M=2.9, SD=0.62), likely due to its technical limitations. While acknowledging that self-reported ratings may be biased by novelty effects, we take this data as a reference rather than validation. Instead, we discuss more details on participants' design experiences based on their feedback, comments, and expectations collected from the study observation and semi-structured interviews.


\subsubsection{Analog Shape-Kit}

When the participants first engaged with analog Shape-Kit during the warm-up, several “\textit{wow!}” moments occurred, especially when they first pressed the pins, saw them transduced to the other side, felt transformed touch on their hands, and realized the system was bi-directional. These reactions highlighted the novel impression of Shape-Kit’s touch design method.

In the post-study interview, participants across all teams agreed crafting haptics with Shape-Kit is intuitive, though some noted how their engagement “\textit{evolved}” (Jamie). 
Mel pointed out the learning curve for handling Shape-Kit, saying, “\textit{I didn’t realize the realm of possibility until the middle part of the study. For instance, using the finger, pen, or phone to actuate the pins in different manners.}” Despite this initial confusion, her team (T2) exemplified collaborative sensorial exploration, resulting in well-executed outcomes such as the Sailboat Dream design.
While several participants noted that the current normal-force-only form factor limited more complex explorations, they all appreciated Shape-Kit’s capability to render a wide range of tactile sensations, especially when incorporating props and materials. 

Participants also emphasized Shape-Kit’s value for rapid prototyping as “\textit{it’s easy, very fast, and you can immediately sense the results}” (Ali). Meanwhile, some appreciated the approach of “\textit{going from the ground up and starting with analog, manual, textural before going to digital. Because even if you can code it in the fanciest equations, it might fail to translate the feeling you want to create}” (Marcelle). The collaborative nature of Shape-Kit was highlighted, as it facilitated idea sharing through touch. Leo pointed out, “\textit{you can kind of only express it through feeling this}” due to the lack of shared vocabulary. Shape-Kit, under this circumstance, provided “\textit{a great way to start the conversation}” (Emilie).

\subsubsection{Touch Pattern Digitization}
During the initial exploration with Shape-Kit, several participants noticed the moving markers in the window module and were curious about their purpose. Aashina exclaimed: “\textit{Oh! You can see that these little guys are moving!}” When we demonstrated the ad-hoc tracking function and the GUI with real-time simulation, participants further expressed their excitement. Arielle remarked: “\textit{It’s so interesting that you can create something physical and have a digital memory of what you’ve created!}” Beyond achieving touch documentation, designers also found the digital simulation of crafted patterns visualized and amplified design details, providing them valuable references to reflect on their creations (see Section \ref{result:bare_hand}). Although the Tuning function in the GUI was not used during the study, participants were intrigued and convinced by its potential.

\subsubsection{Programmable Display Playback}
When testing the playback with the programmable display, participants offered varied feedback. Many commented that while the playback replicated the animated shapes, it could not generate enough pressure. But it worked well for recreating simpler sensations like “\textit{the single motor back and forth one is kinda good (Jamie).}” Participants also frequently noticed “noise” in the motion irregularities during playback. The pins often jiggled as if “\textit{trying very hard}” (Mel) to achieve precise pin displacement at the right moment. These unnatural movements were described as “\textit{robotic} (Emilie), “\textit{choppy}” (Marcelle), or “\textit{itchy}” (Aashina) that led to laughter when trying on foot sole. Interestingly, T5 found that the irregularities added a sense of liveliness to Pet Hug sensation, which was designed for the BOT topic.

Despite these challenges, participants appreciated the tangible playback as an effective demonstration of the full-cycle vision. Arielle remarked: “\textit{it’s a very cool system, especially the fact that you can record and play it back.}” Vivian also said: “\textit{I really see the vision. I think it’s just the servo limitation.}” Participants expressed optimism about the method’s potential, believing it could be significantly enhanced with more advanced technologies, such as those featured in Fig. \ref{fig:advanced-wearable-pin-displays}.

%% file: 6_takeaways_and_insights.tex
\revise{
\section{Key Takeaways on Haptic Toolkit Design}
This section outlines key takeaways on haptic toolkit design, emphasizing how Shape-Kit’s unique form factor supports the collaborative crafting of on-body haptics and how its hybrid functionality fosters exploratory and reflective creation. 

\vspace*{6pt}
\noindent\textbf{Embracing Improvised Props Provokes Open-ended Touch Exploration}

\noindent Drawing from the paintbrush metaphor, which needs pigments to create art, Shape-Kit is intentionally designed to integrate everyday objects and materials as props in the haptic crafting process. These improvised props, with diverse affordances, properties, and textures, expand Shape-Kit’s hand-crafting possibilities and enhance its output's sensory richness. In turn, Shape-Kit transforms static objects and materials with constrained functions into dynamic touch patterns that can actively engage with the human body. This design approach fosters open-ended exploration for more customizable and personal design contexts.

\vspace*{6pt}
\noindent\textbf{Fostering Collaborative Bodily Sensorial Exploration}

\noindent Although touch exploration is often considered intimate, Shape-Kit’s long transducer supports appropriate proxemics between the touch crafter and receiver, respecting personal boundaries in collaborative touch design. The otherness of pin-filtered hand touch creates sensations that feel neutral and non-intrusive, untethered from human-touch metaphors. This encourages exploration beyond social norms while leaving room for alternative interpretations. Shape-Kit’s bi-directional feedback fosters embodied collaboration, enabling the touch crafter and receiver to communicate ideas and refine sensations directly through touch.

\vspace*{6pt}
\noindent\textbf{Analog Crafting Enhances Rapid Prototyping and Nuanced Touch Design}

\noindent 
Crafting with Shape-Kit provides an intuitive entry point for touch design, supporting fast prototyping and rapid testing across body locations and among team members. While crafting novel touch patterns may require specialized training, our findings suggest it excels at uncovering how the subtleties of touch affect the felt quality. This makes it particularly suitable for designing touch patterns that are contextually pleasant and emotionally resonant in everyday scenarios. Despite being limited to normal-force pin actuation, Shape-Kit’s versatility lies in the nuanced variations achieved through diverse crafting methods, body placements, and contextual explorations.

\vspace*{6pt}
\noindent\textbf{Hybrid Design Method Enables Exploratory and Reflective Creation}

\noindent 
As a hybrid design toolkit, Shape-Kit combines analog crafting with touch digitization. By treating touch as design material, crafting haptics emphasizes first-person touch perception, allowing designers to directly sense and understand haptic features through hands-on exploration. Shape-Kit’s digitization feature allows designers to document, visualize, and revisit their crafted patterns. This process goes beyond playback, supporting the review and analysis of design components and decisions from a third-person perspective, further enhancing reflective creation.
}

%% file: 7_discussion.tex

\section{Discussion}

\revise{Reflecting on how designers practiced the novel “crafting haptics” design method enabled by Shape-Kit, two central themes emerged: crafting haptics involves inherent constraints that can also present opportunities, and the collaborative analog touch crafting fosters shared sense-making and deepens the resonance of touch experiences. In this section, we discuss these themes in depth, offering insights into their implications for future haptic design practices.}

\subsection{Crafting Haptics with Constraints}
Our exploration was motivated by introducing an \revise{intuitive} design method for haptics inspired by classical analog drawing, where the human skin serves as the canvas, Shape-Kit functions like a paintbrush, \revise{and props act as pigments. While it is simple to start with, we recognized that designing touch with Shape-Kit is also a process of crafting within constraints:} constraints in form factor, hand control technique, and the knowledge of “how to touch well.” Though appearing as limitations for future improvement, we view those constraints as strengths and opportunities for exploring expressive haptics.

\textbf{\textit{Constraints in Form Factor: }} \revise{During the study, some participants expressed a desire for broader crafting possibilities, such as creating more complex patterns or exploring entirely different haptic modalities. However, the simplicity of the current Shape-Kit, a 5x5 pin array transmitting normal force through 2.5D shape-change, not only aligns with many existing haptic pin displays (Fig. \ref{fig:advanced-wearable-pin-displays}) but also provides an effective starting point. This constraint offers two key advantages: (1) It increases the potential for crafted patterns to be adapted to existing high-tech interfaces. (2) It encouraged designers to focus on iterating and refining specific touch patterns deliberately within a manageable design space. This balance supports rich exploration while reducing the risk of becoming overwhelmed or distracted by unrestricted options.}

\textbf{\textit{Constraints in Designers' Touching Skills and Knowledge:}} Although participants lacked the specialized hand control techniques seen in musicians or bodyworkers, they effectively used props by leveraging their affordances, properties, and textures to craft versatile haptic cues. While designers may not be typically trained in how to “touch well,” human beings have an inherent capacity to sense and interpret touch based on life experiences. This innate capability, combined with an exploratory approach facilitated by tools like Shape-Kit, suggests that designers, even without formal training, can craft high-quality touch sensations by drawing from personal and shared design sensibilities.

These insights point to the broader potential of extending the “crafting haptics” metaphor to other modalities. For instance, materials with varying thermal conductivities could be used to craft thermal haptic feedback. By balancing the proper constraints for each exploration, future work could focus on designing analog, exploratory tools that facilitate collaborative and embodied haptic crafting across different sensory modalities. This approach could help expand the reach of haptic interaction within the design community, encouraging designers to engage with these tactile components and explore more nuanced, multisensory experiences.

\subsection{Crafting Haptics for Resonance}
In crafting expressive haptics with Shape-Kit, we observed a phenomenon we term “crafting resonance.” This concept captures the continuous and concurrent sense-sharing within the design team, encompassing both the touch-feeling and meaning-making levels.

\textit{\textbf{On the touch feeling level}}, the bi-directional force and tactile feedback from Shape-Kit enabled deep engagement across all roles. Unlike typical Wizard-of-Oz setups \cite{dahlback1993wizard} where “wizards” usually merely mimic automatic machines, the crafter and holder were fully immersed in an embodied dialogue with the perceiver. The crafter could continuously feel dynamic resistance from the springs, materials, and the perceiver’s skin and muscles, which informed real-time fine-tuning of the touch actuation. The perceiver, focused on interpreting the sensations, could press specific pins on the output side to signal desired locations and force levels, guiding the crafter to refine the pattern with precision. The holder, meanwhile, not only ensured proper placement on the body but also sensed the force and intensity of the touch patterns, adjusting their grip to maintain optimal contact. Together, they crafted a resonance that arrived at the intended touch experience.

This collaborative, improvisational exploration echoes practices from bodystorming and movement-based design \cite{schleicher2010bodystorming, loke2013strange, fdili2019making, vega2023resources}, where role-playing and in-situ “acting it out” foster embodied cognition. Shape-Kit extends these concepts into dynamic on-body haptic design, with bi-directional tactile feedback provoking implicit tension and negotiation between designers and the tool, as well as among the designers themselves. It surpasses existing haptic “sketching” materials \cite{windlin2022sketching, sondergaard2020designing}, which often rely on one-way interactions targeting the perceiver’s first-person experience, by enabling the entire design team to bring their somatic sensibilities into the process concurrently, fostering deeper somaesthetic resonance.

\textit{\textbf{On the meaning-making level}}, touch is inherently subjective, and the challenge lies in ensuring that sensations created by one person resonate with others, even if not universally \cite{Strahl2021validity}. When a design metaphor was familiar or easily imaginable to the entire team (e.g., alarm clock sound, sea wave, or a hug), or when it emerged from sensorial exploration proposed by one designer and was quickly associated by others (e.g., Morning Dew and Wakeup Stretch in T6), collaborative discussion and sensorial exploration allowed them to prototype, iterate, and refine the designs to achieve the intended sensation. 

However, when ideas were rooted in one designer’s personal history or memories, which often served as a critical source of inspiration, that designer would guide the team in exploring a feeling that matched their memory. Examples include Yifan’s Cat Kneading, Vivian’s memory of Fish Biting, and Arielle’s imagined Bed Drop. This process required sharing and convincing others of how the sensation represented the intended meaning, sometimes leading to disagreements yet often introducing the team to new touch memories.

In these instances, participants frequently noted the absence of a shared verbal language to describe the target feeling, relying instead on crafting the sensation to “speak for itself.” Shape-Kit played a crucial role in facilitating the sharing of these experiential qualities \cite{hook2015somaesthetic}, helping the team reach a shared understanding and building a new form of resonance in meaningful touch design through physical experience.

%% file: 8_limitations_and_future_potentials.tex

\revise{
\section{Limitations and Future Potentials}
Through our study, we identified several limitations of the current Shape-Kit toolkit. In this section, we discuss potential system improvements based on participants’ suggestions and reflections from the research team. Additionally, while our study primarily focused on exploratory and sensorial design, we also reflect on how crafting haptics with Shape-Kit could be applied in practical contexts to better engage with the haptics community in HCI.
\subsection{Toolkit Design and Development}
\subsubsection{Analog Crafting Capability Can Be Enriched}

\begin{figure*}[h]
    \includegraphics[width=\textwidth]{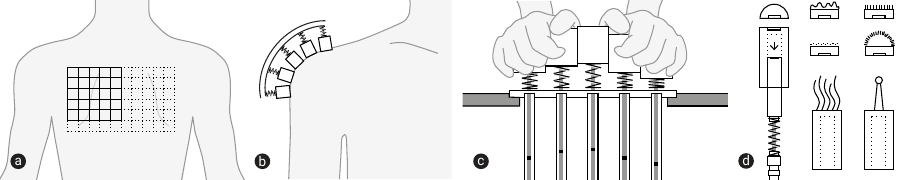}
  \caption{\revise{Participants suggested potential improvements for the analog Shape-Kit (a) Expandable resolution and scale. For instance, a larger pattern size will be necessary for larger body areas like the back. (b) Flexible display frames can better adapt to organic body shapes. (c) Alternative handling form factor. For instance, embedding Shape-Kit on a table-like surface could support crafting like dough-kneading (d) Customizable extensions for individual pins could support richer tactile exploration}}
  \Description{Four sketches to showcase the ideas from the participants to improve the design of analog Shape-Kit. (a) a sketch of a human back where the current shape-kit’s scale and resolution are relatively small, so we usea dashed line to indicate expandable display is needed (b) a sketch of a human shoulder. A flexible shape-kit frame that could make the pin align better to the shoulder shape (c) a sketch of how the hands can do dough kneading gesture well if the shape-kit can be embedded into a table-like surface (d) several example ideas about how the pin cap and tip can be customizable}
  \label{fig:enrich_crafted_touch}
\end{figure*}

While respecting the simplicity of its 2.5D pin-based design, participants suggested several ways to enhance Shape-Kit’s crafting capabilities: 

\noindent• \textit{Expandable Resolution and Scale (Fig. \ref{fig:enrich_crafted_touch}a):} Shape-Kit’s limited resolution and scale posed challenges when crafting meaningful sensations for larger body areas like the back (see Section 5.2.1). This could be addressed by making the display’s resolution and overall scale expandable through modular, joinable displays or an adjustable holding frame.


\noindent• \textit{Flexible Display Frame (Fig. \ref{fig:enrich_crafted_touch}b):} The current flat frame limited its ability to conform to the body’s organic curves. Participants proposed a flexible frame to better adapt to areas like the shoulders for more effective touch delivery. 

\noindent• \textit{Alternative Handling Form Factor (Fig. \ref{fig:enrich_crafted_touch}c):} Designers currently craft sensations by holding the input display by hands, which can be restrictive, especially when using both hands or applying force-intensive gestures like dough-kneading (explored by T1). Embedding the display into a table-like surface could support more dexterous crafting.

\noindent• \textit{Customizable Extensions for Individual Pins (Fig. \ref{fig:enrich_crafted_touch}d):} While various props were explored, most fabric swatches covered the entire display, limiting precise actuation. Participants wanted customizable pin extensions, such as specialized pin caps or tips, to create richer tactile experiences while maintaining precise pin control.

\subsubsection{“Force” Feedback Needs To Be Enhanced}
Many participants reported that the programmable shape display could not provide sufficient force during touch playback. The display in our study was powered by an array of micro linear servos \footnote{AGFRC C1.5CLS-Pro Datasheet: https://m.media-amazon.com/images/I/71zC6HsTupL.pdf}, offering a responsive operation speed (0.108s/9mm@6.0V) and a maximum calculated linear force of 16.43N at 240 g-cm@6.0V torque. However, this peak force was only achievable when the pin first made contact with the skin at full speed. Due to the plastic gears, which are fragile and susceptible to burning out under excessive resistance, participants had to place the display on the body with minimal pressure, unlike the constant pressing possible with the analog Shape-Kit. To enable more effective touch playback, future research should explore more robust haptic actuators capable of applying consistent force feedback when pressing against the skin.
}

\subsubsection{“Force” Digitization Needs To Be Considered}
Our current optical tracking system only captured pin displacement over time without accounting for variations in applied force. As a spring-back system, according to Hooke’s Law \footnote{https://en.wikipedia.org/wiki/Hooke\_law}, pin displacement is linearly related to the force required to move it. However, when pins press into the skin, additional forces are needed to overcome resistance from the skin, muscle tissue, or materials to deliver perceptible force feedback.

To address this, besides integrating force sensors, we propose approximating force data with our current setup. During the documentation process in the study, participants usually actuated the pins without attaching the display to the body. Our findings suggest that crafters sensed the touch pattern through resistance felt in their hands (see Section\ref{sec:property}), indicating that they likely applied similar force levels as when the display was attached to the skin. By recording crafted patterns in both attached and detached conditions and comparing the tracking data through computational models, we could potentially capture approximate force patterns. 

\revise{

\begin{figure*}[h]
  \includegraphics[width=\textwidth]{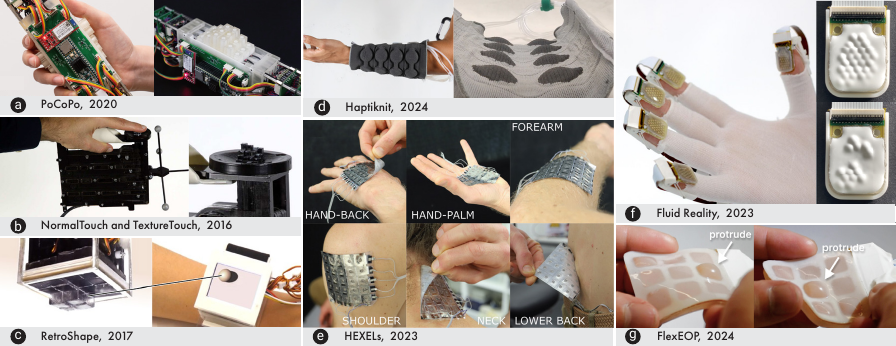}
  \caption{\revise{Selected work on handheld and wearable haptic pin displays. Miniaturized Mechanical Actuators: (a) PoCoPo \cite{yoshida2020pocopo} (b) NormalTouch and TextureTouch \cite{benko2016normaltouch} (c) RetroShape \cite{huang2017retroshape}.
  Pneumatic Actuators: (d) Haptiknit \cite{du2024haptiknit}. Hydraulically Amplified Electrostatic Actuators: (e) Full body HAXELs \cite{leroy2023hydraulically}. Electroosmotic Pump Arrays: (f) Fluid Reality \cite{shen2023fluidreality} (g) FlexEOP \cite{Yu2024FlexEOP}}}
  \Description{A collection figures of existing work on handheld and wearable haptic pin displays. (a) pocopo, a handheld servo-driven pin array (b) normaltouch and texture touch, a handheld device that can render shape-based sensations on the index finger tip (c) retroshape, a watch on the wrist that can deliver pin-based feedback from servo box (d) knit haptic pneumatic sleeve, a knitted sleeve that embedded eight pneumatic bubbles (e) HEXELs, hydraulically amplified electrostatic actuator display that can be applied to hands, arm, knee, neck, and lower back (f) fluid reality, electroosmotic pump array that can deliver rich tactile sensations on finger tip as a haptic glove (g) FlexEOP, electroosmotic pump array with larger pouches and flexible frame}
  \label{fig:advanced-wearable-pin-displays}
\end{figure*}

\subsection{Practical Potentials of Crafting Haptics with Shape-Kit}

Our study prioritized first-person felt experiences of touch, enabling speculative exploration beyond technological constraints. However, while pin arrays are standard in haptic interfaces, their applications are less familiar to the general audience than vibration motors. Shape-Kit’s current displays are also relatively bulky, making it challenging for audiences to envision real-life deployments.

To illustrate the practical potential, Figure \ref{fig:advanced-wearable-pin-displays} presents emerging wearable pin-array hardware technologies recently presented in HCI and Haptics domains, including handheld and wearable devices powered by miniaturized mechanical actuators \cite{yoshida2020pocopo, benko2016normaltouch, huang2017retroshape}, pneumatic actuators \cite{du2024haptiknit}, hydraulically amplified electrostatic actuators \cite{leroy2023hydraulically} and electroosmotic pump actuators \cite{shen2023fluidreality, Yu2024FlexEOP}. These advancements demonstrate how expressive touch crafted with Shape-Kit could be implemented using next-generation actuators in various contexts. While Shape-Kit’s display is rigid, the tracked touch data can be applied to both rigid and non-rigid haptic displays \cite{zhang2022pulling, boem2019non-rigid}, as the key haptic variables are usually the shape, force, and rhythm of the dynamic pin patterns. Moreover, because many soft materials textures were already employed during the Shape-Kit design study, several crafted sensations were perceived as non-rigid. A future direction could explore whether advanced pin displays should continue to incorporate prop materials for tactile enhancement or simulate texture features through specialized algorithms.

We also propose potential form factors for implementing some of the study outcome ideas, drawing inspiration from existing works. While those existing works may not typically employ pin-array displays, they suggest promising directions for integrating expressive haptics. For instance, many SLE patterns crafted during the study could be integrated into furniture and everyday objects \cite{grah2015dorsal, haynes2024breath, hook2016somaesthetic} like mattresses (Bed Drop), pillows (Footsteps, Bug Walking), and blankets (Sailboat Dream). Similarly, wearable devices such as haptic bracelets \cite{young2019bellowband}, sleeves \cite{zhu2020pneusleeve, shtarbanov2023sleevio}, suits \cite{delazio2018forcejacket, TESLASUIT}, clips \cite{choi2022clippable}, or even embedded into fabrics \cite{kilic2021omnifiber} could deliver sensations designed for continuous on-body touch feedback. Many patterns designed for foot sole areas, like Walk on Sand, Morning Dew, and Fish Biting, highlight footwear as another promising application form factor \cite{elvitigala2022ticklefoot}, particularly for the NOI prompt. Additionally, for the BOT topic, haptic patterns could be delivered through embedded pin arrays in stand-alone robots \cite{hu2023skin} or by having virtual robots communicate expressive signals through wearable devices \cite{zhou2020hextouch}. These examples underscore Shape-Kit’s potential not only as a tool for exploratory touch design but also as a bridge connecting speculative haptic design with real-world technological advancements in expressive haptics.

}





%% file: 9_conclusion.tex
\section{Conclusion}

We present Shape-Kit, a hybrid haptic design toolkit that bridges the gap between exploratory design through analog touch prototyping and reproducibility through ad-hoc digitization. It embodies our “crafting haptics” design metaphor, where we envision designing expressive on-body haptics through intuitive hand-crafting, inspired by how visual design stems from analog drawing on canvas with paintbrushes. This paper introduced the motivation and design process behind Shape-Kit. Our design study, involving 14 designers and artists, demonstrated how Shape-Kit facilitated collaborative sensorial exploration and diverse crafting methods to create rich and versatile touch stimulations. \revise{The findings highlight how Shape-Kit's unique form factor and operation method advance expressive haptic design through crafting haptics. While serving as a haptic design tool for sensorial exploration, we also propose how its crafted outcomes could potentially be implemented into practical applications using advanced haptic technologies.}

%% file: main.bbl

%% file: appendix.tex

\appendix
\section{Early Prototype}
\label{section:early_prototype}

\begin{figure}[h]
  \includegraphics[width=0.48\textwidth]{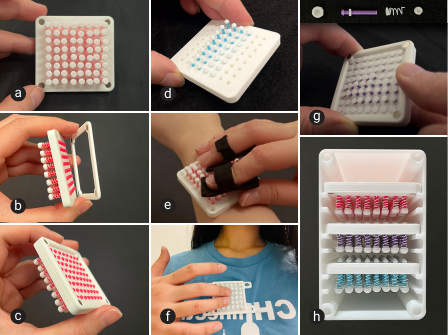}
  \caption{Early Prototype of Shape-Kit}
  \Description{photos show the early prototype of the pin-based touch transducer. These are small spring-back pin displays made with 3D-printed frame, custom springs, and plastic pins. Detailed dimensions and design can be found in the paper}
  \label{fig:early_prototype}
\end{figure}

To design analog haptic shape displays for exploring various pin-based sensations through hand touch, we first developed a series of small-scale displays (Fig. \ref{fig:early_prototype}) inspired by the pin art toy \footnote{https://en.wikipedia.org/wiki/Pin\_art}, which can filter hand pressing into pin shapes. These displays consist of four primary components (Fig. \ref{fig:early_prototype}g top): a pin board (50x50mm), pins (length: 14mm, diameter: 2mm), pin caps (diameter: 4mm), pin tips (Fig. \ref{fig:early_prototype}a,b,c,e - \textit{Round}; \ref{fig:early_prototype}g-  \textit{flat}; \ref{fig:early_prototype}d - \textit{sharp}), and custom springs (2.8mm coil diameter, 2.0mm coil pitch, 5 revolutions, and 0.3mm cantilever). The pin displays were 8x8 with 5mm pin space, aligned with the two-point discrimination threshold for fingers \cite{lederman2009haptic}. The stroke length of the pins was 10mm, referring to the comfort displacement range of human skin \cite{mahmud2010innovative}. All components were 3D printed using FDM printers, except for the pins, which were trimmed from plastic pins in off-the-shelf pin art toys. After testing a range of commercial springs, we decided to use custom springs created with an open-source midair 3D printing generator \footnote{https://makefastworkshop.com/hacks/?p=20181112}, as they were soft yet spongy to allow flexible touch filtering. Additionally, we designed magnetic spacers to ensure the pin tips did not initially contact the skin (Fig. \ref{fig:early_prototype}b,c) and created custom mounts to explore attaching the display to clothing (Fig. \ref{fig:early_prototype}f).

During our preliminary first-person and within-research-team sensorial exploration, several key insights emerged that informed the final form factor of Shape-Kit. First, we observed that the spring-back mechanism could effectively translate hand-touch behaviors into dynamic pin sensations. Different tip shapes of pins were crucial in generating distinct tactile experiences. However, it became apparent that the pin size needed to be enlarged for the sensations to be clearly perceptible on body areas beyond the sensitive hands and face. Additionally, when a person actuated the pins with their fingers while simultaneously trying to perceive the transduced pin sensations on another body part, they found it difficult to focus on the pin sensations. This was because the cues on the finger pads during actuation were more dominant. Even after insulating the fingers with felt caps (Fig. \ref{fig:early_prototype}e), the direct transduction of hand movement to pin movements remained intentional and predictable, leaving less room for people to appreciate the subtle and nuanced touch variations. When attempting to mount the display onto clothing, we found it difficult to ensure it was closely attached to the skin to fully deliver the touch stimulation. Moreover, we noticed that the current pins were too thin and semi-flexible, with large gaps between them, making it difficult to be actuated with movements that included both lateral and normal forces (e.g., rubbing through the pins). Furthermore, to facilitate focused touch perception, we also explored having another person hold the display and actuate the pins. However, this interaction was only suitable for socially comfortable body areas (e.g., arm, hand, and shoulder \cite{suvilehto2015topography}). The thin form factor of the displays meant that the close proximity of the touch crafter’s hand to the perceiver’s body could still feel intrusive, even when the touch was transduced into pin patterns. We used these insights to iterate and refine our design, ultimately arriving at the final Shape-Kit toolkit.

\section{Pilot Study}
\label{section:pilot_study}
\subsection{Study Process}
We conducted two rounds of pilot studies to learn participants’ potential behaviors and preferences, which informed the design and preparation of our exploration sessions. The pilot study began with an introduction to our Shape-Kit toolkit, followed by providing participants with three design topics with prompts \revise{(see Section \ref{section: study_procedure} \& Appendix \ref{section:topics with prompts})}. Participants were asked to design expressive haptic patterns referred to these topics within 60 minutes. After completing each pattern, they were asked to document the sensation using our tracking system and GUI. To understand their design process, we prepared a few interview questions for each pattern, which includes the name, metaphor, context, and design decision, according to prior related design research \cite{zhou2023tactorbots}. After the design exploration, the participants needed to fill out a questionnaire for system evaluation, followed by a semi-structured interview to gain detailed feedback and learn about their design experiences.

\subsection{Takeaways and Iteration}

\subsubsection{Pilot Session One}
The first session included 2 participants (1F, 1M). Which revealed the following takeaways:

\begin{itemize}
    \item Participants primarily engaged with Shape-Kit by “pressing” the pins using normal force with one or multiple fingers. They found the output sensations limited while they felt it challenging to think more “creatively.”
    \item Participants should be encouraged to wear thin, flexible, and comfortable clothing.
    \item The session setup had participants sitting on chairs at a desk, similar to a conventional lab study. This setup led them to primarily explore upper and front body areas (e.g., hands, arms, head), likely due to the sitting posture.
    \item The topic prompts were printed with letter paper and put on the desk, which felt more like exam materials than inspirational prompts.
    \item We conducted interviews about their design process for each created pattern after the entire session. Participants found it difficult to recall details, and it made the session overly long and exhausting, particularly after the 60-minute design exploration.
    \item We needed to enhance the time control and notify the participants during the design exploration.
\end{itemize}

\subsubsection{Pilot Session Two}
To address the issues identified in pilot one, we implemented several changes for the second pilot session. 
(1) Study Preparation: 
To provoke more diverse hand-crafting behaviors, we drew inspiration from the soma design method, which often begins with bodily practices to sensitize the body \cite{hook2018designing}. We provided slime for participants to play with at the start of the study, heightening their awareness of uncommon tactile sensations and enhancing hand dexterity. 

In designing Shape-Kit, we envisioned that designers could employ various crafting strategies, such as using displays to probe or mold other objects or surfaces. However, the original Shape-Kit’s display housings lacked handles, often leading participants to hold them with both hands (especially the smaller displays), leaving only their thumbs available to actuate the pins. To address this, we designed alternative handles (Fig. \ref{fig:analog_system}) that could be swapped with the original holding drawers, offering more flexibility for different handling methods. 

Believing that tools could facilitate more diverse “sculpting” of shape patterns, we gathered everyday objects to serve as props, including plastic balls and plush toys. Recognizing that texture and material are crucial in touch design \cite{karana2016tuning}, particularly for designers and artists interested in orchestrating rich experiences, we also prepared some fabric swatches and custom covers with various tips (Fig. \ref{fig:study setting}a). By attaching these textures to the actuated shape displays, designers could explore a broader range of touch sensations, combining dynamic force feedback with delicate tactile elements. 

(2) Study Setup and Procedure: 
This time, we booked a larger conference room equipped with a table and chairs. However, we let participants freely choose their postures during the design process and reminded them to wear comfortable clothing. The topic prompts were prepared in slide format and presented on a large screen. We adjusted the interview process by having participants answer questions and fill out questionnaires immediately after completing each pattern design. For better time control, we set a digital timer and provided verbal notifications every 20 minutes.

The second round of pilot was conducted with 3 female participants and led to the following key takeaways:

\begin{itemize}
    \item The introduction of slime, handles, and props effectively facilitated more versatile haptic design approaches.
    \item Participants primarily stood when designing, while they occasionally sat to explore body locations such as the foot sole. However, sitting in chairs was still inconvenient for exploring lower body parts.
    \item Conducting interviews after each design idea was distracting. It would be more effective to allow participants to focus solely on touch exploration. Instead, we should do better video documentation of the entire design exploration. \revise{We could later review the recording to extract the design details from the participants' behavior and conversation.}
    \item Topic prompts presented in slide format were not optimal. \revise{As we could only show at most one topic on a page, which constrained the flexibility.}
    \item The use of digital countdowns and verbal notifications from researchers caused anxiety and disrupted the design experience. A more non-disruptive method for time management is needed.
\end{itemize}

\section{Details in Design Study Procedure}
\subsection{Inspirational Topics with Prompts}
\label{section:topics with prompts}

We provided the following topics with prompts to inspire the design exploration, while the designers could pick one or multiple topics to design for.

\textbf{(1) Sleep \& Wake-up enhancement:} Envision a world where your bed, an aircraft seat, or a self-driving car seat can render haptic sensations to support your sleep or wake you up. How do you imagine these sensations enhancing your transition to and from sleep? Consider various expressiveness levels (gentle, urgent, snooze, etc.) and other environments and scenarios where haptic feedback could meaningfully enhance sleep and wake-up experiences.

\textbf{(2) Haptic white noise:} Imagine an office space where, despite being indoors, you can feel the weather, the gentle touch of wind, or natural phenomena like raindrops, ripples, birdsong, leaves rustling, and flowers blooming through subtle haptic feedback. We envision such a “white haptics” experience, which functions as ambient sensory input, could enhance your awareness of time and space and help you stay connected with the outside world while helping you maintain focus. How would you envision these tactile cues feel like?

\textbf{(3) Companion robot’s haptic language:} Imagine a future where companion robots can communicate with you through touch. How would you design this robot’s tactile language to greet you, say goodbye, or capture your attention? The form of the robot can be limitless: it might be stand-alone, table-top, mounted on the body, travel across the body, or even be virtual and can only be seen through XR glasses. It can also have various personalities, such as polite, gentle, shy, or even naughty. Besides visual or auditory cues, how would you envision the robot’s intentional or expressive touch feel like?

\subsection{Prompts in Shape-Kit Toolkit Warming up}
\label{section:prompts in warming up}

After participants started to try operating the Shape-Kit, the facilitator provided prompts to encourage them to explore the broader potential of crafting haptics with this toolkit:

\begin{itemize}
    \item Single Pin - “Can you try pressing a single pin with different rhythms?”
    \item Continuous Pattern - “Can you create a continuous pattern, such as a waveform, with your hands?”
    \item Body Parts - “Why not try it on body parts beyond just your hands?”
    \item Collaboration - “Would you like to try each other’s designs and sensations?”
\end{itemize}

\subsection{Questionnaire in System Evaluation}
\label{section:questionnaire}
Each participant completed a post-study questionnaire to evaluate their experiences with Shape-Kit. The questionnaire included Likert-scale questions, rated from 1 = Strongly disagree to 5 = Strongly agree).

Haptic Design Experiences of Analog Shape-Kit:
\begin{itemize}
    \item Q1: I felt very confident using the system.
    \item Q2: I thought designing haptic experiences with Shape-Kit was intuitive.
    \item Q3: I believe Shape-Kit enables versatile haptic design.
\end{itemize}
Evaluation of Software and Digital Playback:
\begin{itemize}
    \item Q1: I believe the digital simulation on the Shape-Kit GUI can represent our design well.
    \item Q2: I believe the replay of the touch pattern can render our design well.
\end{itemize}

\subsection{Guiding Questions for Semi-structured Interview}
\label{section:interview}
The semi-structured interview followed the questionnaire, allowing participants to first provide detailed qualitative feedback on their design experiences with the Shape-Kit toolkit. We then prompted them to discuss the metaphor of “crafting haptics” and collaborative haptic design, encouraging a generative discussion format.
\begin{itemize}
    \item How was your design experience with Shape-kit, and how do you feel about the \textit{crafting haptics} metaphor?
    \item How was the collaborative design experience for haptics?
\end{itemize}